\documentclass[aps,pra,twocolumn,showpacs,superscriptaddress]{revtex4} 
\usepackage{graphicx}
\usepackage{amsmath}
\usepackage{amssymb}
\usepackage{epstopdf}
\usepackage{amsfonts}
\usepackage{dcolumn}
\usepackage{bigints}
\usepackage{xcolor}
\usepackage{hyperref}
\usepackage[normalem]{ulem}
\begin{document}
\title{
Driven generalized quantum  Rayleigh-van der Pol oscillators: Phase localization and spectral response}
\author{A. J. Sudler}
\address{Homer L. Dodge Department of Physics and Astronomy,
  The University of Oklahoma,
  440 W. Brooks Street,
  Norman,
Oklahoma 73019, USA}
\address{Center for Quantum Research and Technology,
  The University of Oklahoma,
  440 W. Brooks Street,
  Norman,
Oklahoma 73019, USA}
\author{J. Talukdar}
\address{Homer L. Dodge Department of Physics and Astronomy,
  The University of Oklahoma,
  440 W. Brooks Street,
  Norman,
Oklahoma 73019, USA}
\address{Center for Quantum Research and Technology,
  The University of Oklahoma,
  440 W. Brooks Street,
  Norman,
Oklahoma 73019, USA}
\author{D. Blume}
\address{Homer L. Dodge Department of Physics and Astronomy,
  The University of Oklahoma,
  440 W. Brooks Street,
  Norman,
Oklahoma 73019, USA}
\address{Center for Quantum Research and Technology,
  The University of Oklahoma,
  440 W. Brooks Street,
  Norman,
Oklahoma 73019, USA}
\date{\today}

\begin{abstract}
Driven classical self-sustained oscillators have been studied extensively in the context of synchronization. 
Using the master equation, this work considers the classically driven
generalized quantum Rayleigh-van der Pol  oscillator, which is characterized by linear dissipative gain and loss terms as well as three non-linear dissipative terms. Since two of the non-linear terms break the rotational phase space symmetry,
the
Wigner distribution of the quantum mechanical limit cycle state of the undriven system is, in general, not rotationally symmetric. 
The impact of the symmetry-breaking dissipators on the long-time dynamics of the driven system are analyzed as functions of the drive strength and detuning, covering the deep quantum to near-classical regimes. Phase 
localization 
and frequency entrainment, which are required for synchronization, are discussed in detail. We identify a large parameter space where the oscillators exhibit appreciable phase localization but only weak or no entrainment, indicating the absence of synchronization. Several observables are found to exhibit the analog of the celebrated classical Arnold tongue; in some cases, the Arnold tongue is found to be asymmetric with respect to vanishing detuning between the external drive and the natural oscillator frequency. 
\end{abstract}
\maketitle

\section{Introduction}
Self-sustained classical oscillators
do not only contain a damping term but also a term that serves as an energy source.
The competition between the non-linear damping and linear gain
(sometimes also referred to as anti-damping) terms
introduces, in the absence of an external sinusoidal drive,
a limit cycle, i.e., a stable periodic finite-amplitude trajectory
in position-momentum phase space that is approached in the large time limit
regardless of the oscillator's initial conditions. 
The asymptotic finite-amplitude oscillations of self-sustained oscillators underlie a range of phenomena in the 
social sciences, economics, engineering, and the fundamental sciences, including cardiac rhythms,
cell rhythms, and the synchronous blinking of fireflies and clapping of 
audience members~\cite{pikovsky_book,balanov_book}.

In the classical Rayleigh oscillator, which was discussed in 1883 by Strutt and Rayleigh in the context of clocks, violin strings, and clarinet reeds, the non-linear damping is proportional to $\dot{x}^3$ (here, $x$ denotes the dimensionless position and $\dot{x}$ the dimensionless velocity or momentum)~\cite{strutt1883}. In the van der Pol oscillator, in contrast, the non-linear damping term is proportional to $x^2 \dot{x}$; van der Pol and co-workers applied the corresponding equation of motion in 1928 to model the human heart~\cite{vanderpol1,vanderpol2}. We note in passing that the van der Pol oscillator equation can be obtained from the Rayleigh oscillator equation by substitution and subsequent differentiation~\cite{footnote_conversion}.

This work considers the quantum version of a generalized 
non-linear damping term, namely the term
$(\gamma_{2,\text{vdp}} x^2 + \gamma_{2,\text{ray}} \dot{x}^2)\dot{x}$.
Throughout this work, we refer to the oscillator
with equal non-linear position- and momentum-dependent
damping ($\gamma_{2,\text{vdp}}=\gamma_{2,\text{ray}}$) as the
RvdP (Rayleigh-van der Pol) oscillator~\cite{chia2020,arosh2021} and that with $\gamma_{2,\text{vdp}} \ne \gamma_{2,\text{ray}}$ (both coefficients finite) as the generalized RvdP oscillator.
While the classical 
Rayleigh 
($\gamma_{2,\text{vdp}}=0$; $\gamma_{2,\text{ray}}>0$; we use R for Rayleigh throughout), classical 
van der Pol 
($\gamma_{2,\text{vdp}}>0$; $\gamma_{2,\text{ray}}=0$; we use vdP for van der Pol throughout), and classical RvdP
($\gamma_{2,\text{vdp}}=\gamma_{2,\text{ray}}$) oscillators have been investigated extensively,
the quantum version of the paradigmatic rotationally invariant RvdP
oscillator with non-linear damping term proportional to
$(x^2 + \dot{x}^2)\dot{x}$ was first considered in 2013/2014~\cite{lee2013,walter2014}.
Since then, this model has been used to study various aspects of
 synchronization~\cite{chia2020,arosh2021,lee2013,walter2014,hush2015,ameri2015,weiss2017,amitai2017,amitai2018,sonar2018,koppenhofer2019,mok2020,jaseem2020,li2020,eshaqi2020,cabot2021,thomas2022,lorch2017,lee2014,morgan2015}; moreover, its applicability for 
sensing has also been assessed~\cite{dutta2019}.

Quantum versions of the 
R 
and 
vdP
oscillators, both of which possess limit cycles with broken rotational
phase-space symmetry, were considered by Chia {\em{et al.}}~\cite{chia2020} and Arosh {\em{et al.}}~\cite{arosh2021}.
These works presented an analysis of the quantum 
vdP, 
quantum 
R,
and quantum 
RvdP
oscillators and their classical counterparts. The quantum mechanical systems were found to support relaxation oscillations, a key signature of classical non-linear systems~\cite{chia2020}. 
Phase synchronization, which requires phase localization (e.g., non-rotationally symmetric Wigner function) and frequency entrainment (modification of the system's  frequency from the natural harmonic oscillator to the drive frequency), in the presence of a coherent sinusoidal classical drive has been studied extensively for the quantum version of the  
RvdP
oscillator~\cite{chia2020,mok2020,sonar2018,walter2014,dutta2019}. The understanding of driven systems with unequal $\gamma_{2,\text{vdp}}$ and $\gamma_{2,\text{ray}}$, in contrast, 
    is still in its infancy~\cite{chia2020,arosh2021}. 
    Moreover,
    the deep quantum regime in which the system response such as the susceptibility may---extrapolating based on the behavior found for the RvdP oscillator with vanishing linear gain~\cite{dutta2019}---deviate not only quantitatively but 
    also qualitatively from the system response in the classical regime, has not yet been investigated systematically for the generalized RvdP
    oscillator.

Our main conclusions, which are derived by analyzing numerical and perturbative results of the quantum master equation, are as follows:
(i) In all regimes (classical to deep quantum), the phase localization increases for fixed detuning $\Delta$ with increasing drive strength $\Omega$ for all oscillator types considered, including oscillators that may be characterized as being hybrid R--RvdP
oscillators
or hybrid RvdP--vdP
oscillators; this conclusion is---consistent with the conceptual framework of synchronization, which assumes that the drive keeps the amplitude of $\Omega=0$ limit cycle approximately unchanged---restricted to the weak drive strength regime.
(ii) Systems with non-rotationally symmetric dissipators display  in the deep quantum regime behaviors that distinguish them from the RvdP
oscillator, whose dissipators are rotationally symmetric. 
The phase localization, e.g., depends for fixed $\Omega$ on the sign of the detuning $\Delta$ and may vary non-monotonically as $\Delta$ increases or decreases from $0$.
(iii) The power spectrum in frequency space is determined for several drive strengths and detunings. Phase localization is observed over a much larger parameter space than
frequency entrainment. In the deep quantum regime, the spectral response is very broad
and frequency entrainment is either absent or extremely weak.
(iv) 
The modification of the limit cycle amplitude by the external drive, 
relative to that of the drive-free system, is quantified through a deformation parameter $\bar{D}$.
The deformation parameter displays, just as the number of excitations $\bar{N}$ and phase localization measure $\bar{S}_{\text{q}}$, Arnold tongue-like characteristics.
(v) For a large parameter space, we observe phase localization but no frequency entrainment, indicating the
absence of quantum synchronization. We note that many references
(see, e.g., Ref.~\cite{mok2020})
refer to the measure  $S_{\text{q}}$ employed in our work as 
phase synchronization
as opposed to phase localization; we refrain from referring to $S_{\text{q}}$ as phase synchronization
since---in our definition---phase synchronization requires phase locking and frequency entrainment.

The remainder of this paper is organized as follows.
Section~\ref{sec_theory} introduces the master equation, reviews the connection between the master equation and the classical equations of motion, and 
discusses the observables considered in this work.
Results as functions of the detuning and strength of the external drive, and their interpretation, are presented in Sec.~\ref{sec_results}.
Finally, Sec.~\ref{sec_summary} summarizes.
Technical details are relegated to two appendices.

\section{Theoretical framework}
\label{sec_theory}

\subsection{Quantum systems under study: Master equation}
\label{sec_theory_masterequation}

In the laboratory frame (i.e., the ``non-rotating frame"),
the master equation 
for the density matrix $\hat{\rho}$ of the generalized quantum 
RvdP
oscillator 
in scaled dimensionless units reads~\cite{arosh2021}
\begin{eqnarray}
\label{eq_master}
\dot{\hat{\rho}}=
-\imath [\hat{H}, \hat{\rho}] +
\gamma_1^+ \hat{{\cal{D}}}[\hat{a}^{\dagger}](\hat{\rho}) +
\gamma_1^- \hat{{\cal{D}}}[\hat{a}](\hat{\rho}) + \nonumber \\
\alpha \hat{{\cal{D}}}[\hat{a}\hat{a}](\hat{\rho}) +
\beta \hat{{\cal{D}}}[\hat{x}\hat{a}](\hat{\rho}) +
\delta \hat{{\cal{D}}}[\hat{p}\hat{a}](\hat{\rho}) ,
\end{eqnarray}
where 
the operators $\hat{a}$ and $\hat{a}^{\dagger}$ are the bosonic 
annihilation and creation operators,
$\hat{a}|n \rangle=\sqrt{n} | n-1 \rangle$,
$\hat{a}^{\dagger}|n \rangle=\sqrt{n+1} | n+1 \rangle$, 
$\hat{a}^{\dagger} \hat{a} | n \rangle = n | n \rangle$, 
\begin{eqnarray}
\label{eq_convert1}
\hat{x} = \frac{1}{\sqrt{2}} (\hat{a} + \hat{a}^{\dagger})
\end{eqnarray}
and
\begin{eqnarray}
\label{eq_convert2}
\hat{p} = \frac{1}{\sqrt{2} \imath} (\hat{a} -  \hat{a}^{\dagger}).
\end{eqnarray}
The Hamiltonian $\hat{H}$,
\begin{eqnarray}
\hat{H}=
\hat{H}_0 + \hat{V}_{\text{drive}},
\end{eqnarray}
contains the dimensionless one-dimensional harmonic oscillator Hamiltonian $\hat{H}_0$,
\begin{eqnarray}
\hat{H}_0=\hat{a}^{\dagger} \hat{a}
\end{eqnarray}
(for convenience, the ground state energy is chosen to be equal to $0$),
as well as the external drive $\hat{V}_{\text{drive}}$,
\begin{eqnarray}
\label{eq_drive_full_lab}
\hat{V}_{\text{drive}}=  
\Omega  
\sin
(\omega_D t 
)\frac{\hat{a} + \hat{a}^{\dagger}}{\sqrt{2}}
.
\end{eqnarray}
Here, $\Omega$ (which is assumed to be real) denotes the strength of the drive. 
Equation~(\ref{eq_drive_full_lab}) contains co- and counter-rotating terms.
Neglecting the counter-rotating terms,
the drive within the RWA 
simplifies to
\begin{eqnarray}
\label{eq_drive_RWA}
(\hat{V}_{\text{drive}})_{\text{RWA}}
=
\frac{\Omega}{2 \imath \sqrt{2}}
\left(
e^{\imath \omega_D t } \hat{a} -
e^{-\imath \omega_D t } \hat{a} ^{\dagger}
\right).
\end{eqnarray}
Looking ahead, we define the detuning $\Delta$ between the external drive
and the
natural angular
frequency of the harmonic oscillator Hamiltonian $\hat{H}_0$ (in our case, this angular frequency is equal to one),
\begin{eqnarray}
\Delta=\omega_D-1.
\end{eqnarray}

The master equation, Eq.~(\ref{eq_master}), contains five dissipators
$\hat{{\cal{D}}}$, which---for
 an arbitrary operator $\hat{C}$---are defined through
\begin{eqnarray}
\hat{{\cal{D}}}[\hat{C}](\hat{\rho})=
\hat{C} \hat{\rho} \hat{C}^{\dagger}
- \frac{1}{2} 
\{ \hat{C}^{\dagger} \hat{C},\hat{\rho} \},
\end{eqnarray}
where $\{ \hat{A}, \hat{B} \}$
denotes the anti-commutator between the operators $\hat{A}$ and $\hat{B}$.
The dissipators that are proportional to the coefficients $\gamma_1^+$ and $\gamma_1^-$ represent incoherent
linear or one-excitation processes while those that are proportional to
$\alpha$, $\beta$, and $\delta$ represent
incoherent non-linear or two-excitation processes.
Specifically,
$\gamma_1^+$ and $\gamma_1^-$
are incoherent linear
gain and incoherent linear damping rates, respectively. 
The dissipator proportional to the incoherent two-photon damping coefficient $\alpha$ appears in various contexts including phonon lasers and lasing~\cite{simaan1975,dodonov1997,li2020}.
The  terms that are proportional to $\beta$ and $\delta$, in contrast, are studied comparatively rarely~\cite{chia2020,arosh2021}. The physical interpretation of these terms is relegated to
Sec.~\ref{sec_theory_classical}, which connects the quantum equations of motion to the
classical equations of motion for the generalized RvdP
oscillator.


\begin{widetext}

In the eigen basis $\{ | k \rangle \}$
of $\hat{H}_0$, the master equation reads
\begin{eqnarray}
\label{eq_densitymatrix_diff}
\dot{\rho}_{k,l} =
\nonumber \\
-
\imath \frac{\Omega}{\sqrt{2}}\sin(\omega_D t)
\left(
 \sqrt{k} {\rho}_{k-1,l} + 
 \sqrt{k+1} {\rho}_{k+1,l} - 
 \sqrt{l+1} {\rho}_{k,l+1} - 
 \sqrt{l} {\rho}_{k,l-1}  \right)
\nonumber \\
+\left[
-\imath (k-l)
-\frac{\gamma_1^+}{2}(k+l+2)
-\frac{\gamma_1^-}{2}(k+l)
-\frac{\alpha}{2} \left(k(k-1)+l(l-1) \right)
+\left( \frac{\beta}{4} + \frac{\delta}{4} \right)
(2k l -2 k^2 -2 l^2 + k +l)
\right] \rho_{k,l} \nonumber \\
+ \gamma_1^+ \sqrt{kl} \rho_{k-1,l-1}
+ \gamma_1^- \sqrt{(k+1)(l+1)} \rho_{k+1,l+1}
+\left( \alpha + \frac{\beta}{2} + \frac{\delta}{2} \right) \sqrt{(k+1)(k+2)(l+1)(l+2)} \rho_{k+2,l+2} \nonumber \\
+\frac{1}{4} (\beta - \delta)
\Big[ (2l-k) \sqrt{(k+1)(k+2)} \rho_{k+2,l}
+(2k-l) \sqrt{(l+1)(l+2)} \rho_{k,l+2} \nonumber \\
+(2-l) \sqrt{l(l-1)} \rho_{k,l-2}
+(2-k) \sqrt{k(k-1)} \rho_{k-2,l} \Big],
\end{eqnarray}
\end{widetext}
where we introduced the notation
$\rho_{k,l}=\langle k|\hat{\rho}|l \rangle$
and
$\dot{\rho}_{k,l}=\langle k|\dot{\hat{\rho}}|l \rangle$.
To understand the system dynamics, it is useful to summarize the structure of the coupled differential equations. 
For $\Omega=0$, Eq.~(\ref{eq_densitymatrix_diff}) shows the following:
\begin{itemize}
    \item $\beta=\delta $~\cite{footnote1}:
    The coupled differential equations for $\dot{\rho}_{k,l}$ decouple into $N$ independent sets of equations; specifically, 
    the equation for $\dot{\rho}_{0,0}$ is only coupled to the equations for $\dot{\rho}_{n,n}$, where $n=1,\cdots,N-1$;
    the equation for $\dot{\rho}_{0,1}$ is only coupled to the equations for $\dot{\rho}_{n,n+1}$, where $n=1,\cdots,N-2$; and so on.
    Since the coherences can be shown to decay to zero in the large time limit, 
    the stationary $\Omega=0$ limit cycle state is characterized by $\rho_{k,l}=0$ for $k \ne l$, i.e., it is diagonal in the energy eigen basis of $\hat{H}_0$. 
    Reference~\cite{simaan1975} provides analytical expressions for the $\rho_{k,k}$ 
   that characterize the limit cycle. A diagonal density matrix yields a rotationally invariant Wigner function $W(x,p,t)$ (see, e.g., Refs.~\cite{arosh2021,dutta2019}), i.e., a Wigner function that
   depends on $r$ but not on $\varphi$; here, $x=r \cos \varphi$ and $p =r \sin \varphi$. 
\item $\beta \ne \delta$:
    The coupled differential equations for the $\dot{\rho}_{k,l}$
    can be divided into two sets,
    one set for  $\dot{\rho}_{k,l}$ with $k-l$ even and another set for $\dot{\rho}_{k,l}$ with $k-l$ odd. The stationary $\Omega=0$ limit cycle state is characterized by $\rho_{k,l}=0$ for all odd $k-l$. Owing to the dissipators that are proportional to $\beta$ and $\delta$, the limit cycle state is
    non-diagonal in the energy eigen basis of $\hat{H}_0$. Correspondingly, the Wigner function depends explicitly on $r$ and $\varphi$.
\end{itemize}
The drive (i.e., a finite $\Omega$) introduces coupling between $\dot{\rho}_{k,l}$ equations with $k-l$ even and $k-l$ odd.
Specifically, in the weak driving limit, a perturbative order-by-order treatment
(using the framework discussed in Ref.~\cite{koppenhofer2019})
shows that the drive introduces terms that are proportional to $\Omega$ in the off-diagonals $\rho_{k,k \pm 1}$ when $\beta=\delta$
and in the elements $\rho_{k,l}$ with $k-l$ odd when $\beta \ne \delta$.


Throughout, we are interested in parameter combinations
for which the density matrix at large times reaches a state  that displays regular oscillations around a quasi-stationary state. 
Our simulations prepare the 
system at $t=0$ in the coherent state 
$|\alpha_0 \rangle$~\cite{quantum_optics_book}.
The time evolution of the matrix elements $\rho_{k,l}$
is determined by solving the set of first-order coupled differential equations 
given by Eq.~(\ref{eq_densitymatrix_diff}).

We visualize the system at time $t$ using the
phase space Wigner function $W(x,p,t)$,
\begin{eqnarray}
\label{eq_wigner_def}
W(x,p,t) = \frac{1}{\pi} \int_{-\infty}^{\infty} 
\langle x+y | \hat{\rho} | x-y \rangle
e^{-2 \imath p y} dy
,
\end{eqnarray}
which is a quasi-probability~\cite{quantum_optics_book}. 
A key feature of the Wigner function 
is that it connects naturally with the classical phase space trajectories of the corresponding classical system.

\subsection{Connection with classical equation of motion}

\label{sec_theory_classical}

Starting with
 the equation of motion
for $\langle \hat{a} \rangle$,
\begin{eqnarray}
\label{eq_eom_quantum}
\frac{d \langle \hat{a} \rangle}{dt}=
\mbox{Tr} \left( \dot{\hat{\rho}} \hat{a} \right),
\end{eqnarray}
our goal is to obtain an approximate differential equation for
$\langle \hat{x} \rangle$
that maps---in the limit that $\epsilon$, 
\begin{eqnarray}
\epsilon=\gamma_1^+-\gamma_1^-,
\end{eqnarray}
is small---to the classical equations of motion for the generalized driven RvdP
oscillator.
Positive and negative $\epsilon$ correspond to 
net linear gain and
net linear damping, respectively. Importantly, when establishing the mapping between the quantum and classical equations of motions, only the difference between $\gamma_1^+$ and $\gamma_1^-$ enters and not the actual values~\cite{arosh2021}. In the quantum regime, the system characteristics have, however,  been shown to depend on the actual values of $\gamma_1^+$ and $\gamma_1^-$~\cite{arosh2021}. 
Expectation values are, as usual, calculated via the trace operation; e.g., $\langle \hat{a} \rangle = \text{Tr}(\hat{\rho}\hat{a})$.
Introducing
$\bar{\alpha}=\alpha/\epsilon$,
$\bar{\beta}=\beta/\epsilon$, 
$\bar{\delta}=\delta/\epsilon$,
and
$\bar{\Omega}=\Omega/\epsilon$ as well as replacing terms like
$\langle \hat{a}^{\dagger}\hat{a}\hat{a}\rangle$
by
$\langle \hat{a}^{\dagger}\rangle\langle\hat{a}\rangle\langle\hat{a}\rangle$, 
one derives---generalizing the steps of Arosh {\em{et al.}}~\cite{arosh2021}---at order $\epsilon$:
\begin{eqnarray}
\label{eq_quantum_eom_epsilon_main}
\frac{d^2 \langle \hat{x} \rangle}{dt^2}
+ \langle \hat{x} \rangle
= 
-
\epsilon \bar{\Omega} \sin ( \omega_D t) +
\nonumber \\
\epsilon \left[
1 - ( \bar{\alpha}+2 \bar{\beta}  - \bar{\delta} ) \langle \hat{x} \rangle^2
-(\bar{\alpha} + \bar{\delta})
\left( \frac{d \langle \hat{x} \rangle}{dt} \right)^2 \right]\frac{d \langle \hat{x} \rangle}{dt}.
\end{eqnarray}
Physically, the approximations amount to assuming that the classical limit cycle is only weakly deformed and that quantum fluctuations are small. The quantum-classical correspondence for the strongly non-linear regime (lifting of the former restriction) can be established by allowing for additional terms in $\hat{H}_0$. Reference ~\cite{chia2020} carried such a program out for the R,
RvdP,
and vdP oscillators.

Defining
$\bar{\gamma}_{2,\text{vdp}}=\bar{\alpha}+2\bar{\beta}-\bar{\delta}$
and
$\bar{\gamma}_{2,\text{ray}}=\bar{\alpha}+\bar{\delta}$ and replacing the expectation value $\langle \hat{x} \rangle$ by the classical variable $x(t)$, 
Eq.~(\ref{eq_quantum_eom_epsilon_main})
can be identified as 
the classical equation of motion for a driven self-sustained oscillator:
 \begin{eqnarray}
    \label{eq_classical_eom}
    \ddot{x}(t) + x(t)
     = 
-
\epsilon \bar{\Omega} \sin((1+\Delta)t) +
    \nonumber \\ 
    \epsilon \left[ 1 - \bar{\gamma}_{2,\text{vdp}}   (x(t))^2 -
    \bar{\gamma}_{2,\text{ray}} (\dot{x}(t))^2
    \right] \dot{x}(t)  .
    \end{eqnarray}
The subscript ``2" 
reflects that $\bar{\gamma}_{2,\text{vdp}}$
and
$\bar{\gamma}_{2,\text{ray}}$
characterize non-linear damping 
processes;
throughout, these coefficients are assumed to be greater than or equal to zero.
The subscripts ``vdp" and ``ray"
stand for ``van der Pol" and ``Rayleigh," respectively.
The combinations
$(\bar{\gamma}_{2,\text{vdp}}>0,\bar{\gamma}_{2,\text{ray}}=0)$ and
$(\bar{\gamma}_{2,\text{vdp}}=0,\bar{\gamma}_{2,\text{ray}}>0)$
correspond to the paradigmatic
vdP
and 
R oscillators~\cite{chia2020,arosh2021}. The case where the damping rates
$\bar{\gamma}_{2,\text{vdp}}$ and $\bar{\gamma}_{2,\text{ray}}$
are equal is referred to as the 
RvdP
oscillator~\cite{chia2020,arosh2021}.
In terms of $\bar{\alpha}$, $\bar{\beta}$, and $\bar{\gamma}$, the three special cases are:
vdP oscillator with 
$\bar{\alpha}=-\bar{\delta}$,
RvdP
oscillator with
$\bar{\beta}=\bar{\delta}$,
and 
R
oscillator 
with $\bar{\alpha}+2\bar{\beta}=\bar{\delta}$.
 By changing $\bar{\gamma}_{2,\text{vdp}}$ and $\bar{\gamma}_{2,\text{ray}}$ continuously one can  tune from one oscillator type to another.

\begin{widetext}

\begin{figure}[t]
\vspace*{-0.4in}
\includegraphics[width=0.85\textwidth]{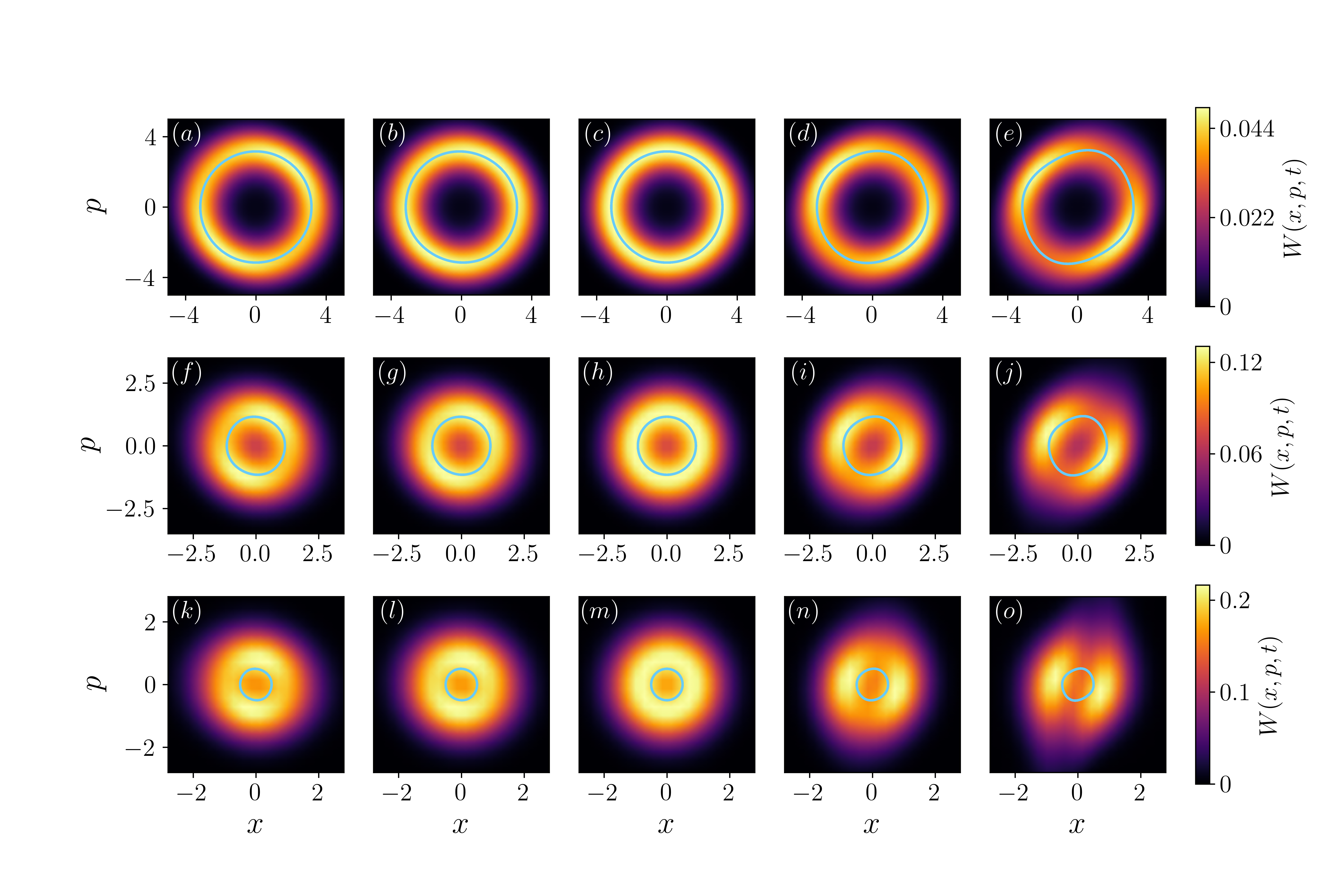}
\vspace*{-0.3in}
\caption{Stationary quantum Wigner functions
$W(x,p,t)$ [Eq.~(\ref{eq_wigner_def})] as functions of $x$ and $p$ in the absence of an external drive; the calculations are performed in the laboratory frame using Eq.~(\ref{eq_densitymatrix_diff}) with $\Omega=0$.
The linear gain rate $\gamma_1^+$ and linear damping rate $\gamma_1^-$ are equal to $1/5$ and $0$, respectively, yielding $\epsilon=1/5$; note, to amplify the rotational asymmetry, this value of $\epsilon$ is twice as large as that used in all other figures of this paper.
The non-linear damping parameters as well as the values of $\langle \hat{a}^{\dagger} \hat{a} \rangle$ and $A_{\text{lc}}$ ($A_{\text{lc}}$ is the same for all five panels in a given row) can be found in Table~\ref{table_parameters_fig1}.
From top to bottom, the quantumness, 
as measured by $\langle \hat{a}^{\dagger} \hat{a} \rangle$, decreases; specifically,
$\langle \hat{a}^{\dagger} \hat{a} \rangle$
is
$ 5.5-5.6$,
$1.18-1.34$, and
$ 0.53-0.83$ 
for the first, second, and third row, respectively. 
The color scales on the far right apply to all five distributions in the respective row.
Note that the scales of the axes are the same for all panels in a given row but that the $x$- and $p$-ranges decrease from the first to the second and from the second to the third rows.
The first, third, and fifth columns are for the 
R,  RvdP, and vdP oscillators, respectively.
The second column is for an oscillator that is ``in between" the R and Rvdp oscillators while the fourth column is for an oscillator that is ``in between" the RvdP and vdP oscillators. 
The light blue solid lines show the classical limit cycle trajectory.
}
\label{fig1_quantum}
\end{figure}

\end{widetext}

\begin{widetext}

 \begin{table}
\begin{tabular}{c|lc|lc|lc|lc|lc||c|c}
 &  & R &&  R / RvdP && RvdP && RvdP / vdP && vdP & $A_{\text{lc}}$ for Fig.~\ref{fig1_quantum} & $A_{\text{lc}}$ for Figs.~\ref{fig_synchronization}-\ref{figure_dbar} \\ \hline
$\alpha$              & (a) &  $0$     & (b) &  $1/100$ & (c) &  $1/50$ & (d) &  $1/100$ & (e)&  $0$ & ``classical"& ``classical" \\
$\beta$           & & $1/75$ && $1/150$ && $0$ && $1/50$ && $1/25$ &$A_{\text{lc}}=\sqrt{10}$&$A_{\text{lc}}=\sqrt{5}$\\
$\delta$ && $2/75$  && $1/75$ && $0$ && $0$ && $0$ &$\approx 3.162$&$\approx 2.236$\\ \hline
$\alpha$              & (f) & $0$     & (g) & $3/40$ & (h) & $3/20$ & (i) & $3/40$ & (j) & $0$ &``transition"&``transition"\\
$\beta$  && $1/10$  && $1/20$ && $0$ && $3/20$ && $3/10$ &$A_{\text{lc}}=\sqrt{4/3}$&$A_{\text{lc}}=\sqrt{2/3}$\\
$\delta$              && $1/5$   && $1/10$ && $0$ && $0$ && $0$ &$\approx 1.155$&$\approx 0.816$\\ \hline
$\alpha$ & (k) & $0$     & (l) & $2/5$ & (m) & $4/5$ & (n) & $2/5$ & (o) & $0$ & ``quantum"& ``quantum"\\
$\beta$ && $8/15$  && $4/15$ && $0$ && $4/5$ && $8/5$ &$A_{\text{lc}}=1/2$&$A_{\text{lc}}=\sqrt{1/8}$\\     
$\delta $&& $16/15$ && $8/15$ && $0$ && $0$ && $0$ &$= 0.5$&$\approx 0.354$\\ 
\end{tabular}
\caption{The first six columns provide a summary of non-linear damping  parameters
$\alpha$, $\beta$, and $\delta$ used in Figs.~\ref{fig1_quantum}, \ref{fig_synchronization},  \ref{figure_adaggera}, and \ref{figure_dbar}; the labels (a)-(o) refer to the
panels of the figures. For reference, the last two columns report the limit cycle amplitude $A_{\text{lc}}$, which depends on the 
parameter $\epsilon$; $\epsilon=1/5$ and $1/10$ for Fig.~\ref{fig1_quantum} and Figs.~\ref{fig_synchronization}-\ref{figure_dbar},  respectively.
In each figure, the classical limit cycle amplitude $A_{\text{lc}}$ is the same across each row [e.g., panels (a)-(e)].
The labels ``classical," ``transition," and ``quantum"
refer to the quantumness of the system; 
the labels are meant to serve as a rough guide.
}
\label{table_parameters_fig1}
 \end{table}

\end{widetext}


The classical equations of motion can be analyzed using
 secular perturbation theory, in which $\epsilon$ is---consistent with the discussion above---treated as a 
 small parameter~\cite{burton1984,bender_book}.
Defining the scaled detuning
$\bar{\Delta}$ as well as the slow time scale $T$
[$\bar{\Delta}=\Delta / \epsilon$ and
 $T= \epsilon t$;
since $T$ depends on $t$, we use
the notation $T(t)$], one
makes the ansatz
\begin{eqnarray}
\label{eq_ansatz}
x(t) = \frac{1}{2}A(T(t))
\exp( \imath t) + 
\frac{1}{2}
[A(T(t))]^*
\exp(- \imath t) + \nonumber \\
\epsilon x_1(t),
\end{eqnarray}
where the amplitude $A$ is assumed to be a slowly varying function in $T$.
In the absence of an external drive,
the amplitude
$|A(T(t))|=A_{\text{lc}}$,
\begin{eqnarray}
\label{eq_amplitude_lc}
A_{\text{lc}} =2\left(\bar{\gamma}_{2,\text{vdp}}
+ 3 \bar{\gamma}_{2,\text{ray}} \right)^{-1/2}
,
\end{eqnarray}
corresponds to a stable limit cycle (hence the subscript ``lc").
Regardless of where the classical trajectory is started, it approaches at long times 
a trajectory that is, to leading order, characterized by $A_{\text{lc}}$~\cite{pikovsky_book,bender_book}.
The solid light blue lines in Fig.~\ref{fig1_quantum} show the numerically determined classical limit cycle trajectories for
$\Omega=0$ and $\epsilon = 1/5$ for various oscillator types and various non-linear parameters, i.e., various $A_{\text{lc}}$ values.
The limit cycle is a prerequisite
for the emergence of classical  synchronization in the presence of a non-vanishing drive  with $\Omega$ and $|\Delta|$ not too large~\cite{pikovsky_book}.
One objective of the present work is to study the quantum analog of the celebrated classical Arnold tongues, which play a fundamental role in synchronization studies, for the generalized
RvdP
oscillator with rotational phase space asymmetry. 

The mapping between the classical and quantum equations of motion 
motivates the functional forms of the ``non-linear dissipators." Specifically, the arguments of the dissipators that are proportional to $\beta$ and $\delta$ are chosen such that the quantum equations of motion map, for small $\epsilon$, to the classical equations of motion for the R
and 
vdP
oscillators. While the dissipator $\hat{{\cal{D}}}[\hat{a}^2](\hat{\rho})$ has a clear physical interpretation (two-photon losses), it was suggested that an experimental realization of the dissipators $\hat{{\cal{D}}}[\hat{x}\hat{a}](\hat{\rho})$ 
and $\hat{{\cal{D}}}[\hat{p}\hat{a}](\hat{\rho})$ 
may involve measurement and feedback processes~\cite{chia2020}. 

We emphasize that the mapping between the classical and quantum equations of motion is derived in the laboratory frame.
This is important since the terms 
$\hat{{\cal{D}}}[\hat{x}\hat{a}](\hat{\rho})$
and
$\hat{{\cal{D}}}[\hat{p}\hat{a}](\hat{\rho})$
are not invariant under a transformation to a rotating frame (see Appendix~\ref{appendix_rotating}), i.e., the functional form of the master equation changes unless $\beta$ and $\delta$ are equal to each other. 
The terms
$\hat{{\cal{D}}}[\hat{a}^{\dagger}](\hat{\rho})$,
$\hat{{\cal{D}}}[\hat{a}](\hat{\rho})$,
and
$\hat{{\cal{D}}}[\hat{a} \hat{a}](\hat{\rho})$, in contrast, are unchanged under a transformation to a rotating frame.
It follows that
only the master equation for the 
RvdP
oscillator is invariant under a transformation to a rotating frame. In general, a transformation to a rotating frame introduces new terms, which can be interpreted as being due to fictitious forces that arise in response to the 
rotation, in analogy to, e.g., the Coriolis force in classical mechanics~\cite{goldstein_book}.

\subsection{Observables}
\label{sec_theory_observables}

As alluded to earlier,
the expectation value $\langle \hat{a}^{\dagger} \hat{a} \rangle$ measures  the quantumness of the oscillator. The self-sustained oscillator is, in the absence of the external drive, in the classical regime, the crossover regime, and the quantum regime when 
 $\langle \hat{a}^{\dagger} \hat{a} \rangle \gg 1$,
 $\langle \hat{a}^{\dagger} \hat{a} \rangle \approx 1$, and $\langle \hat{a}^{\dagger} \hat{a} \rangle \ll 1$, respectively. Being in the quantum regime requires that the dissipators that lead to a lowering of the excitations are sufficiently strong.
 For the RvdP oscillator
 with $\beta=\delta$, e.g., the effects of the $\alpha$ and $\gamma_1^-$ terms need to dominate over the $\gamma_1^+$ term.
 A non-vanishing external drive tends to act as an energy source, leading to an increase of
$\langle \hat{a}^{\dagger} \hat{a} \rangle$  in the quasi-stationary regime
compared to the situation where the drive strength is zero.
Quite generally, the system dynamics can be divided into two regimes: initial transient dynamics and long-time quasi-stationary dynamics.
Figures~\ref{figure_timetrace}
and \ref{figure_wigner_drive} show the time dependence of observables, covering the transient and quasi-stationary regimes, while Figs.~\ref{fig1_quantum}, \ref{fig_synchronization}, \ref{figure_adaggera},
and \ref{figure_dbar} display the system characteristics in the quasi-stationary regime,
which is---for the parameters considered---governed, at least to leading order, by the limit cycle of the system without external drive.

Our primary interest in this work lies in quantifying phase localization and frequency entrainment, in the non-transient quasi-stationary regime.
Quantum phase localization has been quantified through various measures, including  phase distribution, ``moments" such as $\langle \hat{a} \rangle$, entanglement, and information theory based 
observables~\cite{tindall2020,thomas2022,ameri2015,jaseem2020,mari2013,lee2014}.
Measures that involve
the quantum mechanical phase operator $\hat{\varphi}$~\cite{pegg1988,shapiro1991}
are intuitively appealing as they provide an immediate link to one of the classical 
phase localization
metrics, namely the mean resultant length $S_{\text{cl}}$, which is defined as
$S_{\text{cl}}=\sqrt{ \langle \sin \varphi \rangle^2 + \langle \cos \varphi \rangle^2}=|\langle \exp(\imath \varphi) \rangle |$. In this classical context, the $\langle \cdot \rangle$ notation indicates an ensemble average as opposed to the quantum mechanical trace operation.
If the phases $\varphi$, 
$\varphi = \mbox{atan}(p/x)$,  for the driven self-sustained classical oscillator are distributed uniformly, $S_{\text{cl}}$
is equal to zero. For non-uniformly distributed $\varphi$, on the other hand, $S_{\text{cl}}$
is finite but never larger than $1$.
By analogy, quantum phase localization is quantified through (see, e.g., Ref.~\cite{mok2020}) 
\begin{eqnarray}
\label{eq_synchronization_raw}
S_{\text{q}} = |\langle \exp(\imath \hat{\varphi}) \rangle|.
\end{eqnarray}
It follows straightforwardly that $S_{\text{q}}$ lies---just as the classical mean resultant length $S_{\text{cl}}$---between zero and one:
A value of $S_{\text{q}}=0$ indicates the absence of phase localization while a value of $1$ indicates maximal phase localization.

It is useful to 
rewrite $S_{\text{q}}$,  Eq.~(\ref{eq_synchronization_raw}), in terms of the density matrix elements (see Appendix~\ref{appendix_measure}):
\begin{eqnarray}
\label{eq_synchronization}
S_{\text{q}} = 
\left| \sum_{n=1}^{\infty} \rho_{n,n-1} \right|.
\end{eqnarray}
This expression 
highlights that $S_{\text{q}}$ is governed by the coherences of the density matrix.
Using the properties discussed in the context of Eq.~(\ref{eq_densitymatrix_diff}), it can be shown readily that $S_{\text{q}}$ vanishes 
in the $\Omega=0$ and $t \rightarrow \infty$ limits for all oscillator types considered in this work. Correspondingly, the non-vanishing values of $S_{\text{q}}$ observed in this work are introduced by the external drive.
If $\Omega$ is too large, the external drive may reshape the $\Omega=0$ limit cycle so strongly that the oscillator's amplitude changes notably. To
quantify amplitude distortions, we compare the radius $R$ at which the Wigner function 
takes its maximum for finite and vanishing drive strengths.

We note that
 while Ref.~\cite{koppenhofer2019} states that a rotationally invariant limit cycle is a prerequisite for 
 finite phase localization, 
 we quantify phase localization also for $\Omega=0$ limit cycles that possess broken rotational symmetry, i.e., for limit cycles that are characterized by non-zero $\rho_{k,l}$ elements for $|k-l|=2, 4,\cdots$.
Specifically, the next section investigates whether
 self-sustained oscillators with non-rotationally symmetric limit cycles 
(those with 
$\beta \ne \delta$ or, equivalently, $\gamma_{2,\text{vpd}} \ne \gamma_{2,\text{ray}}$) enhance or hinder quantum 
phase localization. 

To quantify frequency entrainment,
 we calculate the power spectrum $S_p(\omega,\tau)$~\cite{walter2014}, 
 \begin{eqnarray}
 \label{eq_powerspectrum}
S_p(\omega,\tau)=
\int_{-\infty} ^{\infty}
C(t,\tau)
\exp(- \imath \omega t) 
dt
,
\end{eqnarray}
which is defined as the Fourier transform of the correlation function
 $C(t,\tau)$,
 \begin{eqnarray}
C(t,\tau)=\langle \hat{a}^{\dagger}(t+\tau)\hat{a}(\tau)\rangle.
\end{eqnarray}
The  correlation function $C(t,\tau)$ is obtained by averaging over the full quantum mechanical density matrix (system and environment), making use of the   regression theorem~\cite{breuer2002}.
Since our main focus lies in characterizing the system behavior in the non-transient quasi-stationary regime, 
we restrict ourselves to sufficiently large $\tau$.
To calculate $S_p(\omega,\tau)$, we work in the frame that
rotates with the drive.
As mentioned earlier and discussed formally in Appendix~\ref{appendix_rotating}, the transformation from the laboratory to the rotating frame changes the functional form
of the dissipators that are proportional to $\beta$ and $\gamma$.
In particular, the dissipators in the rotating frame are, in general, time dependent.
Correspondingly, $S_p(\omega,\tau)$ depends explicitly on $\tau$.
The power spectrum, calculated in the frame rotating at $\omega_D$, 
is expected to exhibit a delta-function like spike at $\omega=0$ as well as 
a broader response, possibly with pronounced side 
peaks~\cite{weiss2016,walter2014,cabot2021,andre2012}.
If the center of a broad, non-delta function-like peak lies
at $\omega=0$ as opposed to at $\omega=\Delta$,
the system is said to be entrained:
the broad response that is associated with the dissipative terms is linked to the drive frequency
 as opposed to the natural harmonic
oscillator frequency. Recall, since the frame is rotating with $\omega_D$,
a response at $\omega=0$ and $\omega=\Delta$ corresponds to being locked to the drive 
frequency and 
to being locked to the natural oscillator frequency, respectively.

\section{Results}
\label{sec_results}

Figure~\ref{fig1_quantum} shows snapshots of quasi-stationary Wigner functions
for the 
undriven generalized 
RvdP
oscillator in the laboratory frame. 
Three regimes are covered:
 the quantum regime
characterized 
by 
$\langle \hat{a}^{\dagger} \hat{a} \rangle < 1 $
(third row), the intermediate regime 
characterized by $\langle \hat{a}^{\dagger} \hat{a} \rangle \approx 1$ (second row), and the 
classical regime
characterized by 
$\langle \hat{a}^{\dagger} \hat{a} \rangle > 1 $ (first row).
In the classical regime, the Wigner function takes on its (local) maxima at  $(x,p)$ values that closely follow the  corresponding classical limit cycle trajectory
(light blue solid line). The close resemblance between the (local) maxima of the quantum mechanical Wigner function $W(x,p,t)$ and the classical limit cycle trajectory in the classical regime confirms the quantum-classical correspondence derived in Sec.~\ref{sec_theory_classical} for the generalized
RvdP
oscillator in the small $\epsilon$ regime (Fig.~\ref{fig1_quantum} employs $\epsilon=1/5$),
thereby extending the work by Arosh {\em{et al.}}~\cite{arosh2021} for the R, RvdP, and vdP
oscillators to include oscillators that lie between the
R and RvdP
oscillators (column~2 of Fig.~\ref{fig1_quantum}) or between the
RvdP and vdP
oscillators (column~4 of Fig.~\ref{fig1_quantum}).
We also checked that classical finite-temperature ensemble calculations, in which the temperature mimicks the role of the quantum fluctuations, reproduces the Wigner functions semi-quantitatively for all oscillator types considered. This observation further 
confirms the  limiting classical behavior derived in Sec.~\ref{sec_theory_classical}.

The quasi-stationary Wigner functions for the 
RvdP
oscillator without external drive (column~3 of Fig.~\ref{fig1_quantum}) are, as discussed in Sec.~\ref{sec_theory_classical}, rotationally symmetric in all regimes (quantum to classical).
For the corresponding classical system, the non-linear damping term (i.e., the term that is proportional to $x^2+\dot{x}^2$) is directly proportional to the  energy and thus constant along the circular trajectory
(solid light blue lines).
As $\alpha$ increases from Fig.~\ref{fig1_quantum}(c)
to Fig.~\ref{fig1_quantum}(h) to Fig.~\ref{fig1_quantum}(m), the non-linear damping becomes stronger, $\langle \hat{a}^{\dagger}\hat{a} \rangle$ 
decreases, and the ring-shaped Wigner function ``shrinks".

The classical limit cycle trajectories for the other oscillator types 
(columns~1, 2, 4, and 5 of Fig.~\ref{fig1_quantum}) are not circular but slightly deformed, reflecting the fact that the non-linear damping terms $x^2\dot{x}$ and $\dot{x}^3$ contribute with unequal strengths. For the vdP oscillator, e.g., the classical trajectory reaches its most positive and most negative $p$-values at finite positive and finite negative $x$-values, respectively (column~5 of Fig.~\ref{fig1_quantum}).
Since the magnitude of the velocity is largest
at these points, the system spends less time in these phase space regions than in others phase space regions.
Interestingly, these classical features are inherited by the quasi-stationary Wigner function, which displays two global maxima that are, roughly, located at $p\approx-x\approx \pm 3$ for the largest
$\langle \hat{a}^{\dagger}\hat{a} \rangle$ considered.
In the quantum regime, the Wigner function of the vdP oscillator is characterized by two ``tilted lobes." While the classical limit cycle trajectory does not capture the detailed structure of the Wigner function, it can be used to estimate the ``tilt angle" and location of the lobe maxima.  
The R
oscillator (column~1 of Fig.~\ref{fig1_quantum}) possesses, as the vdP oscillator, a rotational phase-space asymmetry, with the roles of $x$ and $p$ being reversed in the non-linear damping term. As a consequence, the tilt angle of the 
R 
oscillator differs in the quantum regime by about
$\pi/2$ from that of the  vdP oscillator
[see Fig.~\ref{fig1_quantum}(k)].

\begin{figure}[h]
\includegraphics[width=0.4\textwidth,scale=0.25]{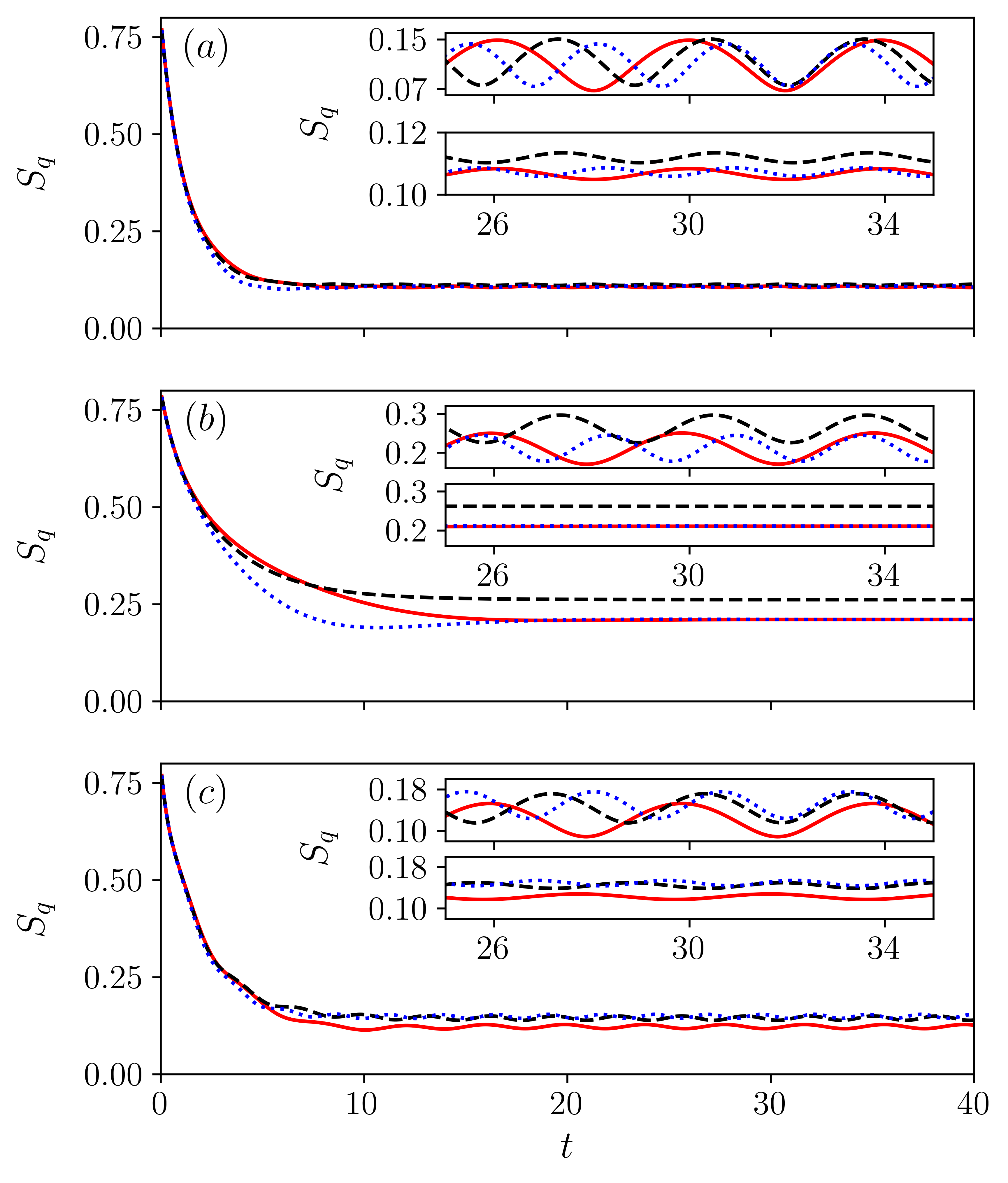}
\caption{Phase localization $S_{\text{q}}$, calculated in the laboratory frame,
as a function of time for 
(a) the 
R
oscillator
[$\alpha=0$, $\beta=8/15$, and $\delta=16/15$; same non-linear parameters as used in Figs.~\ref{fig1_quantum}(k), \ref{figure_wigner_drive}(k), 
\ref{figure_adaggera}(k), and \ref{figure_dbar}(k)], (b) the RvdP oscillator
[$\alpha=4/5$, $\beta=0$, and $\delta=0$; same non-linear parameters as used in Figs.~\ref{fig1_quantum}(m), \ref{figure_wigner_drive}(m), \ref{figure_adaggera}(m), and \ref{figure_dbar}(m)], and 
(c) the vdP oscillator
[$\alpha=0$, $\beta=8/5$, and $\delta=0$; same non-linear parameters as used in Figs.~\ref{fig1_quantum}(o), \ref{figure_wigner_drive}(o), \ref{figure_adaggera}(o), and \ref{figure_dbar}(o)]
with linear parameters
$\gamma_1^+=1/5$ and $\gamma_1^-=1/10$
in the presence of an external drive with $\Omega = 3/10$.
 The black dashed, red solid, and blue dotted 
 lines are for $\Delta = 0, -1/5$, and $1/5$, respectively. 
 The oscillators are, at $t=0$, prepared in a coherent state 
with $x_0=p_0=3/(2 \sqrt{2}) \approx 1.061$.
 The results in the main figure employ the RWA.
 In the quasi-stationary long-time limit ($t \gtrsim 10-20$), the phase localization for the 
 R
 and vdP oscillators display oscillations; these oscillations are enlarged in the lower inset in each of the panels.
 For comparison, the top inset in each of the panels shows the phase localization within the BRWA. Compared to the RWA, the BRWA terms introduce larger amplitude
 oscillations.
 \label{figure_timetrace}
}
\end{figure}

The quasi-stationary Wigner functions, i.e., the quantum mechanical limit cycles, are critical for observing phase synchronization in the presence of an
external drive. Working within the RWA, we consider the regime where the drive is perturbative in the sense that the drive does not destroy the limit cycle that is supported by the system with vanishing drive; this aspect is discussed in more detail below in the context of Fig.~\ref{figure_dbar}.
Figures~\ref{figure_timetrace}(a),
\ref{figure_timetrace}(b), and \ref{figure_timetrace}(c)
show the  phase localization $S_{\text{q}}$,
calculated in the laboratory frame, as a function of time for the
R, RvdP,
and vdP oscillators for
damping and gain parameters that, in the absence of the external drive, fall into the quantum regime.
Since the initial state is a coherent state with relatively well defined phase,
the phase localization decreases approximately monotonically during the transient dynamics ($t \lesssim 10$ in Fig.~\ref{figure_timetrace}) during which the Wigner function moves toward the limit cycle. 
For $t \gtrsim 10$ or $20$,
the phase localization
is essentially constant [Fig.~\ref{figure_timetrace}(b)] or displays regular oscillatory behavior  [Figs.~\ref{figure_timetrace}(a) and \ref{figure_timetrace}(c)]. Figure~\ref{fig_synchronization} includes the transient dynamics to show the order of magnitude of the  time that is needed to reach the quasi-stationary regime. Throughout, we are interested in physics that is independent of the initial state. Because of this we should, strictly speaking, refer to the quantity $S_{\text{q}}$ as phase localization
 only in the quasi-stationary regime ($t \gtrsim 10-20$) and not in the transient regime
 (recall phase localization is
a necessary but not sufficient condition for phase synchronization).

\begin{figure}
\vspace*{-0.0cm}
\includegraphics[width=0.4\textwidth]{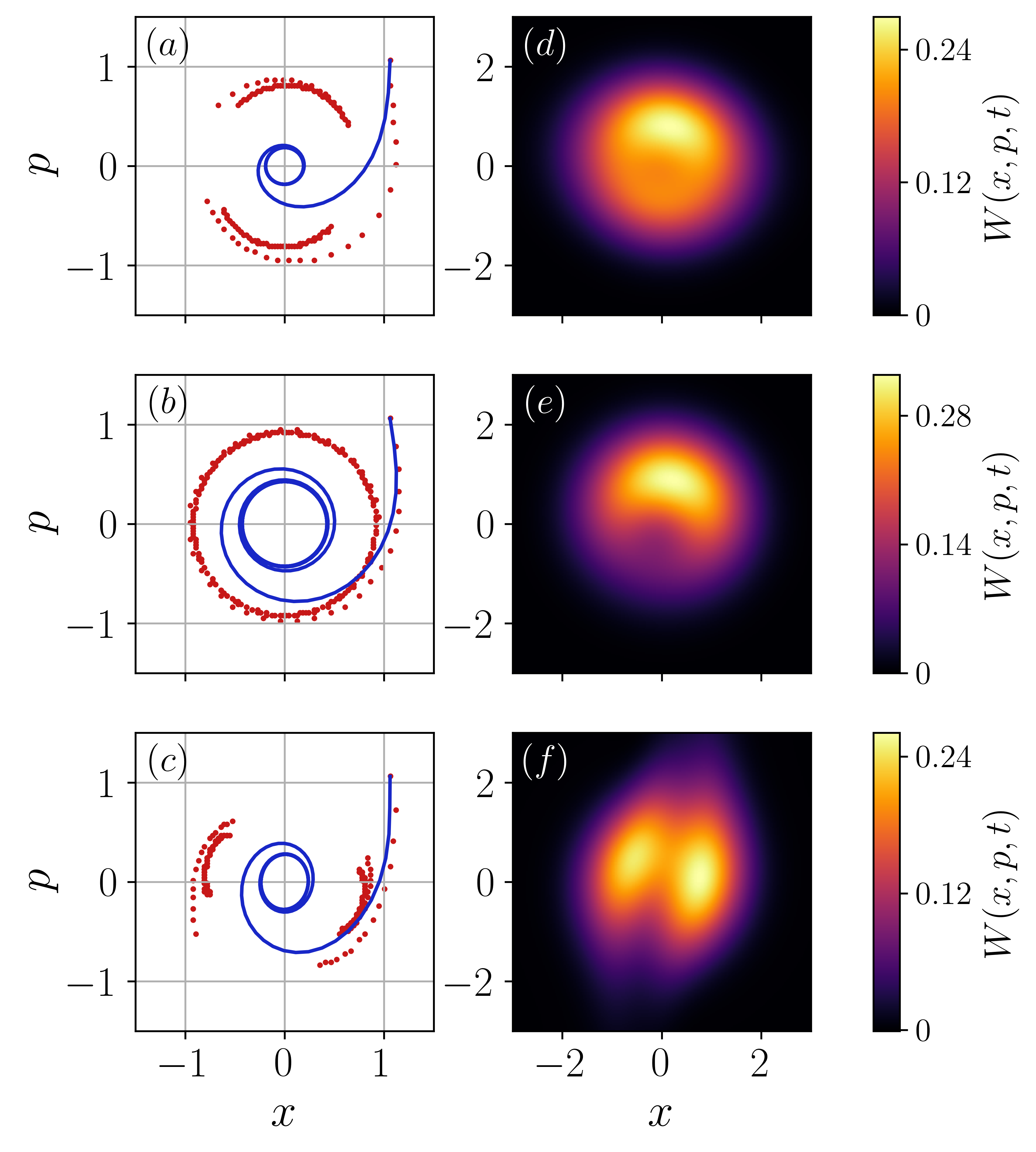}
\vspace*{-0.4cm}
\caption{Dynamics for $0 \le t \lesssim 50$ (left column) 
and Wigner function at $t \approx 30$ (right column) 
in the laboratory frame,
calculated within the RWA, for (a)/(d) the R
oscillator, (b)/(e) the RvdP 
oscillator, and (c)/(f) the vdP
oscillator
in the presence of an external drive with $\Omega = 3/10$ and $\Delta=0$.
The linear and non-linear parameters in (a)/(d), (b)/(e), and (c)/(f) are the same as in Figs.~\ref{figure_timetrace}(a),  \ref{figure_timetrace}(b), and \ref{figure_timetrace}(c), respectively.
The blue line and red dots show the center-of-mass trajectory and the trajectory at which the Wigner function is maximal, respectively;
the red dots are calculated at evenly spaced time intervals.}
\label{figure_wigner_drive}
\end{figure}

\begin{widetext}

\begin{figure}[t]
\includegraphics[width=0.85\textwidth]{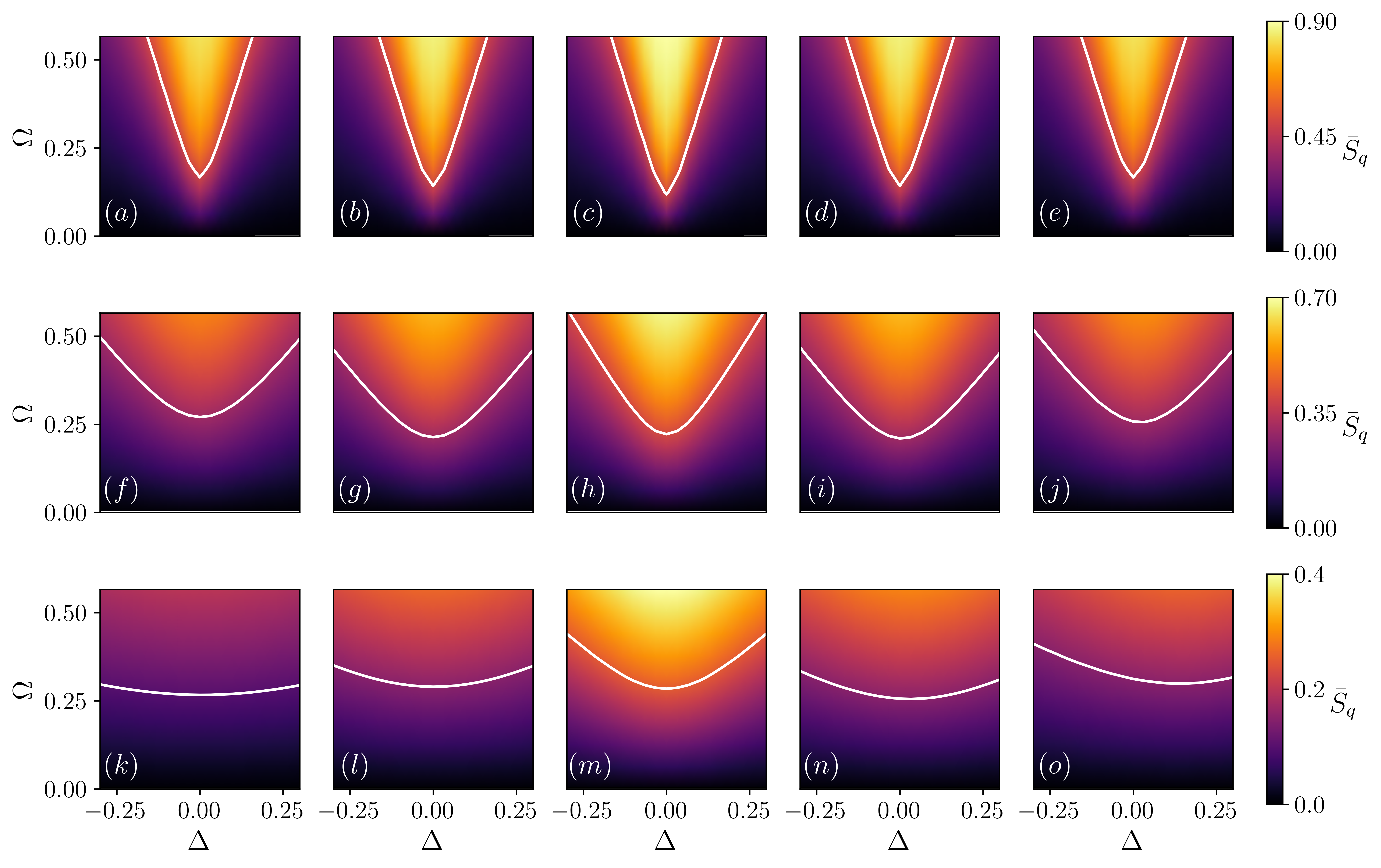}
\caption{Time-averaged phase localization
$\bar{S}_{\text{q}}$ 
as functions of the detuning $\Delta$ and the drive strength $\Omega$; the calculations are performed in the laboratory frame within the RWA.
The linear gain coefficient
$\gamma_1^+$ and linear damping coefficient
$\gamma_1^-$ are equal to $1/5$ and $1/10$, respectively, yielding $\epsilon=1/10$
(this value is smaller than that used in Fig.~\ref{fig1_quantum}). 
The non-linear damping parameters as well as the values of $A_{\text{lc}}$ ($A_{\text{lc}}$ is the same for the five panels in a given row) can be found in Table~\ref{table_parameters_fig1}.
As a guide to the eye, the white lines show 
lines of constant $\bar{S}_{\text{q}}$.}
\label{fig_synchronization} 
\end{figure}

\end{widetext}

\begin{widetext}

\begin{figure}[t]
\includegraphics[width=0.85\textwidth]{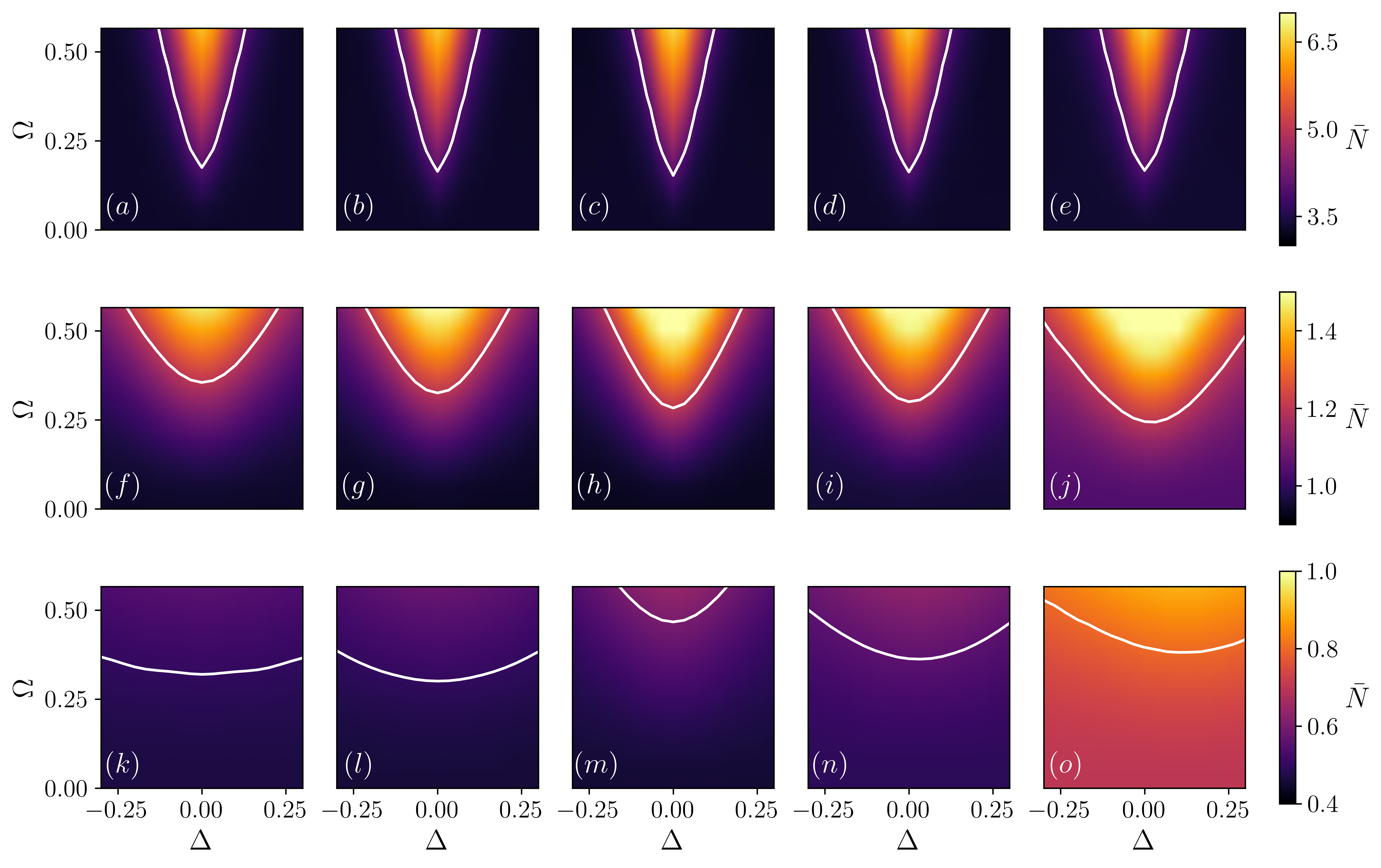}
\caption{Time-averaged number
$\bar{N}$ of excitations
as functions of the detuning $\Delta$ and the drive strength $\Omega$; the calculations are performed in the laboratory frame within the RWA.
The parameters are the same as those used 
in Fig.~\ref{fig_synchronization}.
}
\label{figure_adaggera}
\end{figure}

\end{widetext}

\begin{widetext}

\begin{figure}[t]
\includegraphics[width=0.85\textwidth]{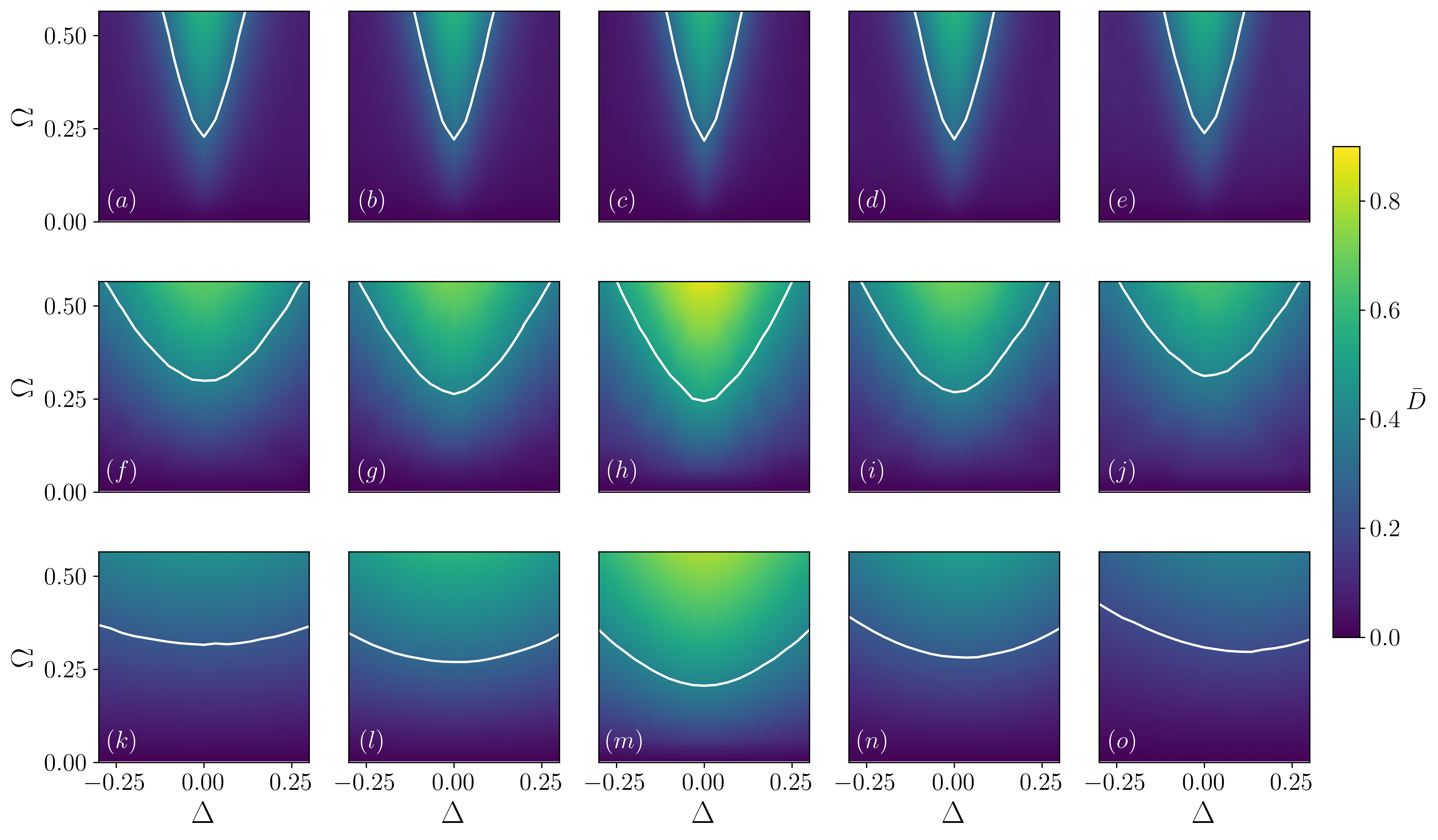}
\caption{Time-averaged deformation
$\bar{D}$ as functions of the detuning $\Delta$ and the drive strength $\Omega$; the calculations are performed in the laboratory frame within the RWA. The parameters are the same as those used in Figs.~\ref{fig_synchronization} and \ref{figure_adaggera}.}
\label{figure_dbar}
\end{figure}    

\end{widetext}

\begin{widetext}

\begin{figure}[t]
\includegraphics[width=0.85\textwidth]{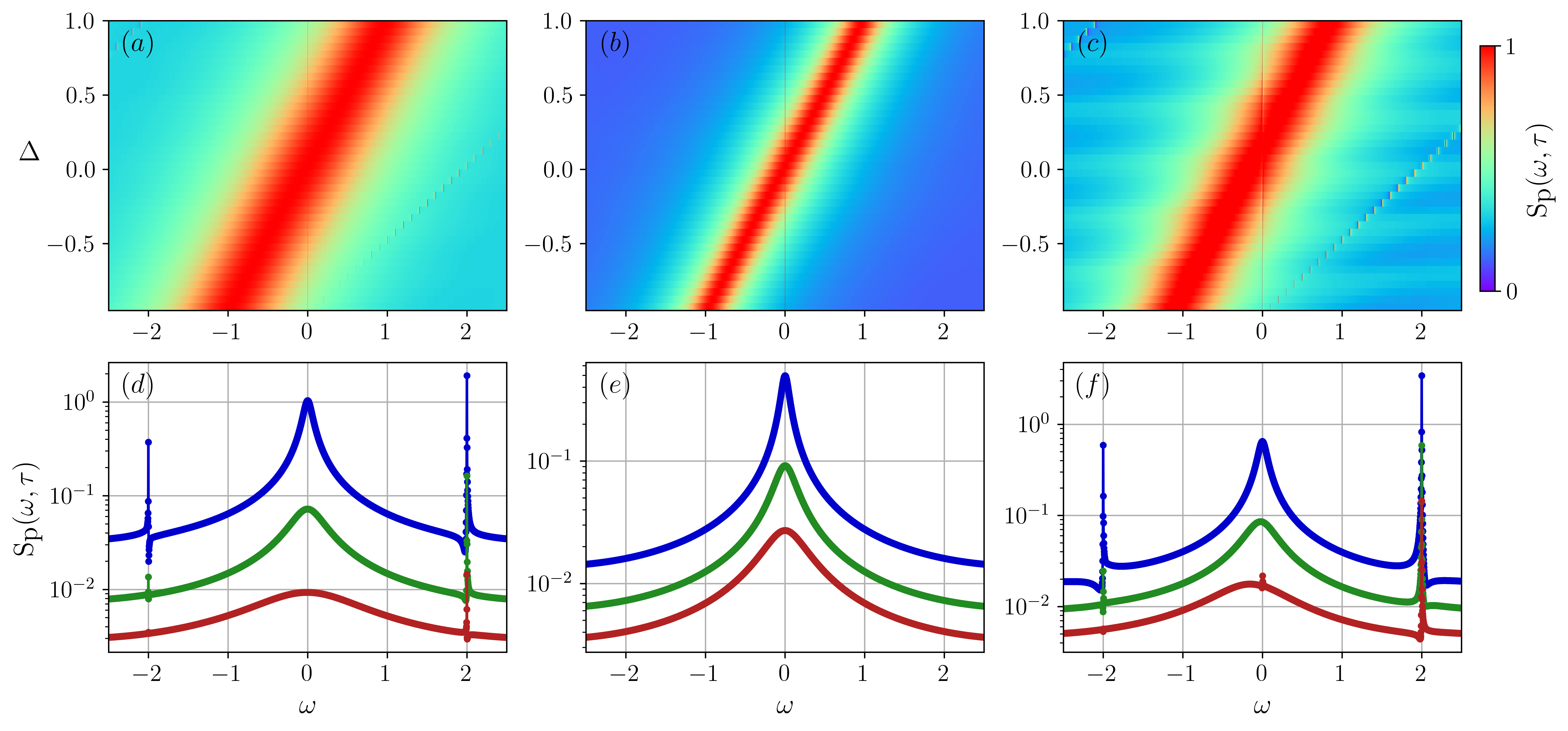}
\caption{The left, middle, and right columns show  power spectra $S_p(\omega,\tau)$ for the R, RvdP, and vdP oscillators, respectively, for $\Omega=3/10$, $\gamma_1^+=1/5$, $\gamma_1^-=1/10$, and $\tau=200$; the  calculations are performed within the RWA in the frame that rotates at the drive frequency. 
 Top row:  Power spectra in the quantum regime as functions of $\omega$ and $\Delta$.
 The non-linear damping parameters for panels~(a), (b), and (c) are given in Table~\ref{table_parameters_fig1}(k), \ref{table_parameters_fig1}(m), and \ref{table_parameters_fig1}(o), respectively.
 For each $\Delta$,
 the power spectrum  is normalized to the maximum of the broad peak that is located at  $\omega \approx \Delta$. The color bar on the right applies to all three spectra.  Bottom row: Power spectra for $\Delta=0$ as a function of $\omega$. The red spectra (bottom curves) are $\Delta=0$ cuts through the data shown in the top row.
 For comparison, the green (middle) and blue (top) curves show power spectra for the transition and classical regimes, respectively.
 The sharp $\delta$-function like data point at $\omega=0$ is not shown and the data are not normalized. Note the log scale of the $y$-axis. 
 For panels~(d)/(e)/(f), the non-linear parameters for the spectra plotted in red (bottom curve), green (middle curve), and  blue (top curve) are given in 
 Table~\ref{table_parameters_fig1}(k)/\ref{table_parameters_fig1}(m)/\ref{table_parameters_fig1}(o), \ref{table_parameters_fig1}(f)/\ref{table_parameters_fig1}(h)/\ref{table_parameters_fig1}(j), and \ref{table_parameters_fig1}(a)/\ref{table_parameters_fig1}(c)/\ref{table_parameters_fig1}(e), 
 respectively.
 }
\label{figure_spectra}
\end{figure}    

\end{widetext}

\begin{widetext}

  \begin{figure}[t]
\includegraphics[width=0.85\textwidth]{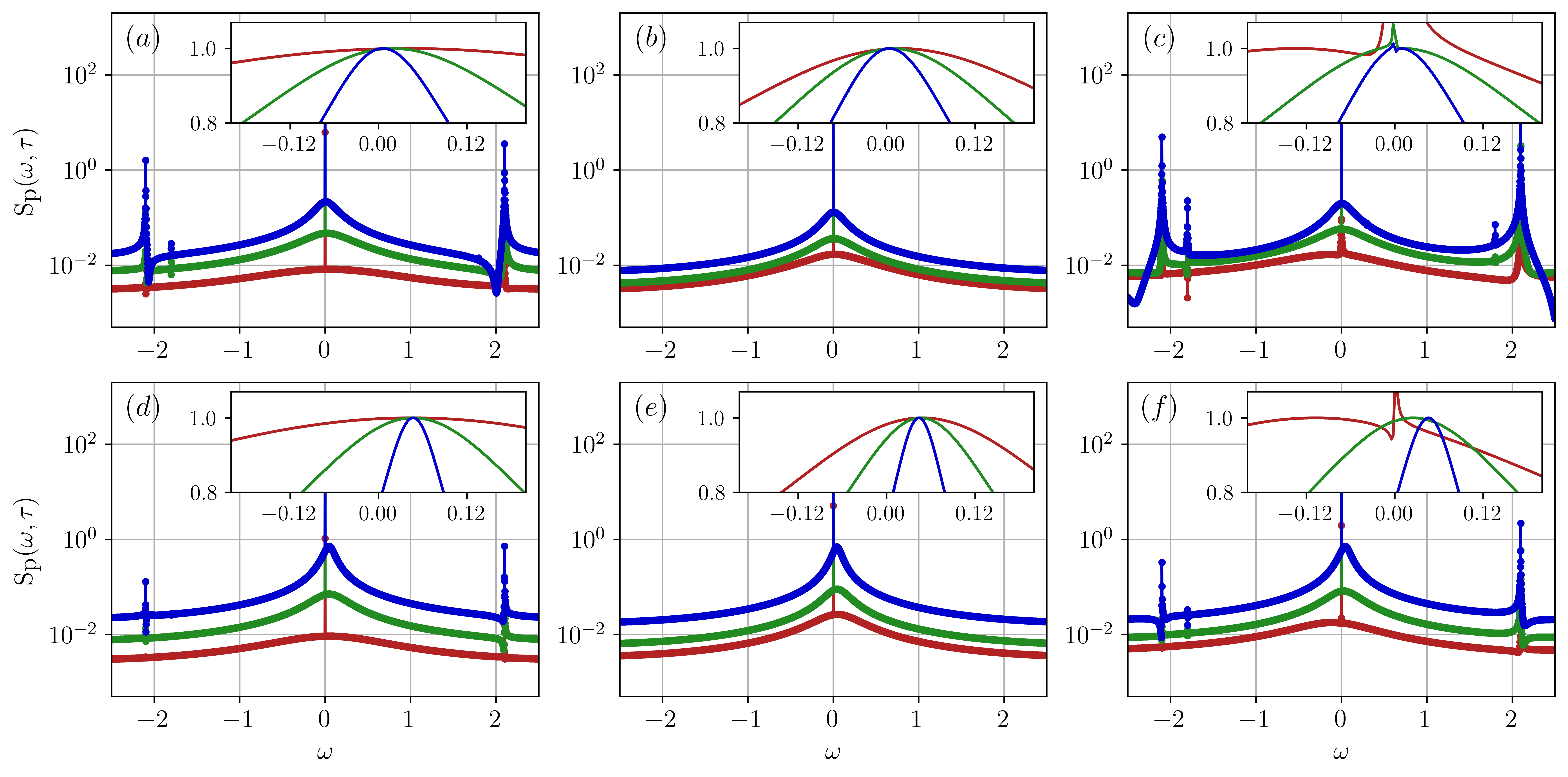}
\caption{The left, middle, and right columns show  power spectra $S_p(\omega,\tau)$ for the R, RvdP, and vdP oscillators, respectively, for $\Delta=1/20$, $\gamma_1^+=1/5$, $\gamma_1^-=1/10$, and $\tau=200$; the  calculations are performed within the RWA in the frame that rotates at the drive frequency. 
 Top row:  Power spectra for $\Omega= 3 \sqrt{2}/5$.
 Bottom row:  Power spectra for $\Omega= \sqrt{2}/5$.
 For panels~(a)\&(d)/(b)\&(e)/(c)\&(f), the non-linear parameters for the spectra plotted in red (bottom curve), green (middle curve), and  blue (top curve) are given in 
 Table~\ref{table_parameters_fig1}(k)/\ref{table_parameters_fig1}(m)/\ref{table_parameters_fig1}(o), \ref{table_parameters_fig1}(f)/\ref{table_parameters_fig1}(h)/\ref{table_parameters_fig1}(j), and \ref{table_parameters_fig1}(a)/\ref{table_parameters_fig1}(c)/\ref{table_parameters_fig1}(e), 
 respectively.
 The insets show blow-ups of the region around $\omega=0$; in these plots, the normalization is 
 chosen such that the maximum of the broad peak is equal to one and the $\delta$-function-like spike is removed. 
 }
\label{figure_spectra2}
\end{figure}    

\end{widetext}

The black dashed, red solid, and blue dotted lines in Fig.~\ref{figure_timetrace}
are for 
$\Delta=0$, $-1/5$, and $1/5$, respectively.
For the 
RvdP oscillator [see Fig.~\ref{figure_timetrace}(b)],
$S_{\text{q}}$ is---in the large time limit---constant (this can be seen in the lower inset of 
Fig.~\ref{figure_timetrace}(b), which shows a blow-up at large times). For the same drive strength, $S_{\text{q}}$ is larger for zero detuning than for finite detuning.
This might be expected naively, as a finite detuning
decreases the ``similarity" of the system and the external drive, thereby hindering phase localization.
In the transient regime, in contrast, $S_\text{q}$
for the RvdP
oscillator depends on the sign of the detuning $\Delta$.
The inclusion of the counter-rotating terms in the external coherent drive leads, as shown in the upper inset of Fig.~\ref{figure_timetrace}(b), to oscillatory behavior of $S_{\text{q}}$ in the  long-time regime. It can be seen that the counter-rotating terms break the symmetry, i.e., the red solid and blue dotted lines (same $|\Delta|$ but opposite signs) are characterized by different oscillation periods 
(the oscillation frequency is equal to $2 \omega_D$) as well as slightly different amplitudes.
As expected for the relatively weak drive strength and small, in magnitude, detuning considered, the counter-rotating terms introduce relatively small corrections. 
Correspondingly,
the results presented in Figs.~\ref{figure_timetrace}-\ref{figure_spectra2} 
of this paper are obtained within the RWA.

Figure~\ref{figure_wigner_drive}(e) shows a snapshot of the Wigner function at $t \approx 30$ for $\Delta=0$ [the other parameters are the same as in Fig.~\ref{figure_timetrace}(b)]. Since $\Omega$ is finite, the Wigner function is not rotationally symmetric but instead displays a half-moon shape. While the shape of $W(x,p,t)$ 
does not change appreciably for $t \gtrsim 10$, the entire distribution rotates with time. This can be seen from the blue line 
in Fig.~\ref{figure_wigner_drive}(b), which shows the quantum mechanical $(\langle \hat{x} \rangle, \langle \hat{p} \rangle)$-trajectory as a function of time. 
In the quasi-stationary regime (i.e., the regime where the shape of the Wigner function does not change), the oscillation frequencies of $\langle \hat{x} \rangle$ and $\langle \hat{p} \rangle$ are regular and equal to $\omega_D$ (which is identical to that of the harmonic oscillator).
For finite $\Delta$ (not shown), the oscillation frequencies of $\langle \hat{x} \rangle$ and $\langle \hat{p} \rangle$ for the 
RvdP oscillator are also equal to $\omega_D$. The red dots in Fig.~\ref{figure_wigner_drive}(b) show the $(\langle \hat{x} \rangle, \langle \hat{p} \rangle)$-values at which the Wigner function is maximal; to make the figure, the Wigner function is analyzed at about 300 equidistantly spaced times between $0$ and $50$.
In the quasi-stationary regime, the red dots trace out a circle.

The phase localization
for the R
and 
vdP
oscillators
[Figs.~\ref{figure_timetrace}(a) and \ref{figure_timetrace}(c)] decreases---similarly to that for the
RvdP oscillator 
[Fig.~\ref{figure_timetrace}(b)]---in the transient short-time regime.
Key differences between the rotationally phase-space asymmetric and rotationally phase-space symmetric oscillators exist, however, in the 
quasi-stationary regime.
For  the 
R
and  
vdP
oscillators, $S_{\text{q}}$---calculated within the RWA---is
 oscillatory with oscillation period $T=\pi/\omega_D$ [see the main panels and lower insets of Figs.~\ref{figure_timetrace}(a) and \ref{figure_timetrace}(c)].  
As expected, inclusion of the counter-rotating terms enhances the oscillation amplitude
[see the upper insets in Figs.~\ref{figure_timetrace}(a) and \ref{figure_timetrace}(c)].
The main panels of Figs.~\ref{figure_timetrace}(a) and \ref{figure_timetrace}(c) show that the  phase localization for oscillator types that have rotationally asymmetric dissipators depends in the
quasi-stationary regime on the sign of the detuning, i.e., the phase localization  displays an asymmetry with respect to $\Delta=0$.

Figure~\ref{figure_wigner_drive} compares the dynamics for the R
oscillator (top row) and the vdP oscillator (bottom row) for $\Delta=0$ (same parameters as used in Fig.~\ref{figure_timetrace}). 
While the $(\langle \hat{x} \rangle, \langle \hat{p} \rangle)$-trajectories (blue lines)  follow a smooth path, the maximum of the
Wigner function (red dots) changes rapidly over a short time interval, leading to a ``bimodal" behavior that is not observed for the RvdP
oscillator
(see middle row of Fig.~\ref{figure_wigner_drive}).
A related bimodality was noted for an oscillator that contains squeezing-like operators~\cite{chia2020}; more specifically,  Ref.~\cite{chia2020} included---to reproduce the classical R oscillator dynamics to higher order in $\epsilon$---terms up to fourth-order in $\hat{a}$ and $\hat{a}^{\dagger}$ in the Hamiltonian and dissipators that are similar in form to our R oscillator. The fact that bi-modality is observed also in our case indicates that the rotational asymmetry of the  dissipators, combined with a rotationally symmetric Hamiltonian, is sufficient for observing bimodality.

Since the 
radius $R$, $R=\sqrt{\langle \hat{x} \rangle^2+ \langle \hat{p} \rangle^2}$, at which the Wigner function is maximal   
is approximately constant for the undriven system in the quasi-stationary long-time regime, we use it to quantify the robustness of the limit cycle 
to the external drive. 
Specifically, 
we define the average deformation $\bar{D}$,
\begin{eqnarray}
\label{eq_deformation}
\bar{D} = 
\lim_{t_{\text{ref}} \rightarrow \infty}
\frac{1}{T} \int_{t_{\text{ref}}}^{t_{\text{ref}}+T}
 \frac{|R_{\text{driven}}(t)-R_{\text{undriven}}|}{R_{\text{undriven}}} 
 dt,
\end{eqnarray}
where  $t_{\text{ref}}$
is chosen such that the system dynamics is in the quasi-stationary regime, $T$ denotes the oscillation period,  and
$R_{\text{undriven}}$ and $R_{\text{driven}}(t)$ refer to the
radii at which the Wigner distribution of the undriven ($\Omega=0$) and driven ($\Omega \ne 0$)  systems is maximal for identical parameters (except 
for $\Omega$).
While the radius $R_{\text{driven}}(t)$ of the driven system depends on time,
the radius $R_{\text{undriven}}$ of the undriven system is, provided $t_{\text{ref}}$ is sufficiently large, independent of time.
 A value of $\bar{D} $
close to zero signals that the drive has a perturbative effect on the amplitude of the limit cycle. 
The larger the value of $\bar{D}$ is, the more the amplitude of the limit cycle is modified by  the external drive  (the limit cycle might even get destroyed).
Recall, the concept of phase synchronization assumes that the external drive
localizes the phase while leaving the amplitude of the limit cycle approximately unchanged.
For the parameters considered in Fig.~\ref{figure_wigner_drive}, $\bar{D}$ is equal to $0.24$ (top row),
 $0.52$ (middle row),
 and
  $0.20$ (bottom row).
Interestingly, for the same drive strength $\Omega$, the  deformation $\bar{D}$ of the RvdP oscillator is larger than the deformation of the R and vdP oscillators
(see also Fig.~\ref{figure_dbar}).

Figure~\ref{fig_synchronization} reports the time average $\bar{S}_{\text{q}}$,
\begin{eqnarray}
\label{eq_sync_ave}
\bar{S}_{\text{q}}
= \lim_{t_{\text{ref}} \rightarrow \infty} \frac{1}{T}
\int_{t_{\text{ref}}}^{t_{\text{ref}}+T} S_{\text{q}}(t) dt
\end{eqnarray}
[$S_{\text{q}}(t)$ is calculated within the RWA], as functions of $\Omega$ and  $\Delta$ for 15 
parameter combinations
(the non-linear parameters are provided in Table~\ref{table_parameters_fig1}).
Since $S_{\text{q}}(t)$ does not oscillate for the
RvdP
oscillator 
[see Fig.~\ref{figure_timetrace}(b)], we set $\bar{S}_{\text{q}}=S_{\text{q}}$ in this case.
The first, third, and fifth columns of Fig.~\ref{fig_synchronization} 
are for the R, RvdP, and vdP
oscillators. The second and fourth columns are for oscillators that lie ``in between" those featured in the neighboring columns.
The top, middle, and bottom rows are for parameters that fall, roughly, into the classical, transition, and quantum regimes.

Figure~\ref{fig_synchronization} shows that, for the drive strengths and detunings considered, the maximum of the phase localization $\bar{S}_{\text{q}}$ for each of the five  oscillator types is larger in the classical regime than in the quantum regime. 
 It is attributed to the fact that the extent or size of the Wigner functions 
decreases with decreasing $\langle \hat{a}^{\dagger}\hat{a} \rangle$
while the  fluctuations (or ``fuzziness") increases~\cite{walter2014,ameri2015,lee2013a}.
The average phase localization
$\bar{S}_{\text{q}}$ for the 
RvdP
oscillator (third column of Fig.~\ref{fig_synchronization}) displays, in agreement with what was found 
previously~\cite{mok2020}, 
the celebrated Arnold tongue behavior.
Specifically, the average phase localization $\bar{S}_{\text{q}}$ is symmetric with respect to $\Delta=0$, increases (for the parameter combinations considered) with increasing $\Omega$ for fixed $\Delta$, and generally decreases with increasing $|\Delta|$ for fixed $\Omega$. Comparing
Figs.~\ref{fig_synchronization}(c), \ref{fig_synchronization}(h), and \ref{fig_synchronization}(m), 
it can be seen that the tongue becomes broader in the deep quantum regime, i.e., the relative change with $|\Delta|$ for
fixed $\Omega$ is  smaller in the deep quantum regime than in the classical regime.

 Comparing the values of the time-averaged phase localization
$\bar{S}_{\text{q}}$ for the different oscillator types with comparable $\langle \hat{a}^{\dagger}\hat{a} \rangle$, i.e., across rows,
it can be seen that the maximum of the phase localization decreases as the dissipators that are not rotationally phase-space symmetric are turned on and play an increasingly important role.
While the average phase localization $\bar{S}_{\text{q}}$  
for the oscillators with rotationally phase-space asymmetric dissipators (first, second, fourth, and fifth columns in 
Fig.~\ref{fig_synchronization}) behaves---in the classical and transition regime---rather 
similarly to that for the
RvdP
oscillator, clear differences are
apparent in the deep quantum regime. Specifically, Figs.~\ref{fig_synchronization}(k), \ref{fig_synchronization}(l),
\ref{fig_synchronization}(n), and \ref{fig_synchronization}(o) reveal the following:
(i) The average phase localization $\bar{S}_{\text{q}}$ is not symmetric with respect to $\Delta=0$;
(ii) starting at the $\Delta$ value for which $\bar{S}_{\text{q}}$ is minimal for a given $\Omega$,  $\bar{S}_{\text{q}}$ 
does not necessarily increase monotonically as one moves along the $\Delta$ axis; and
(iii) for the parameters considered, $\bar{S}_{\text{q}}$ depends less strongly on $\Delta$
in  Figs.~\ref{fig_synchronization}(k), \ref{fig_synchronization}(l),
\ref{fig_synchronization}(n), and \ref{fig_synchronization}(o) than in Fig.~\ref{fig_synchronization}(m).

Figure~\ref{figure_adaggera} shows the time-averaged number $\bar{N}$ of excitations  for the same 15 parameter combinations as those used in Fig.~\ref{fig_synchronization}.
Similar to $\bar{S}_{\text{q}}$,
 $\bar{N}$ is calculated by averaging 
 $\langle \hat{a}^{\dagger} \hat{a} \rangle$  in the quasi-stationary regime over one time period. A visual comparison of
 $\bar{N}$ and $\bar{S}_{\text{q}}$ reveals a striking similarity of the two observables. While the black color represents different background values (zero in the case of $\bar{S}_{\text{q}}$ and a non-zero value in the case of $\bar{N}$), 
$\bar{S}_{\text{q}}$ and $\bar{N}$ appear to be changing in a  correlated manner. Specifically, the
time-averaged number $\bar{N}$ of excitations, plotted as functions of the detuning and drive strength, exhibits---in the classical and transition regimes---Arnold tongue-type characteristics.
The fact that the observables
$\bar{S}_{\text{q}}$ and $\bar{N}$ display similar characteristics, when visualized in terms of color plots as functions of $\Delta$ and $\Omega$, may---at first sight---seem surprising as these two observables depend on different density matrix elements.
As shown in Eq.~(\ref{eq_synchronization}),
$S_{\text{q}}$ is governed by the off-diagonal density matrix elements $\rho_{n,n-1}$;
$\langle \hat{a}^{\dagger} \hat{a} \rangle$, in contrast, is governed by the diagonal density matrix elements $\rho_{n,n}$.
Since the off-diagonal elements of the density matrix, which determine the phase localization, are within first-order perturbation theory 
proportional to $\Omega$ (see Appendix~\ref{appendix_measure}), with a proportionality factor that depends on 
the $\rho_{n,n}$ and $\rho_{n,n+2}$ ($n=1,2,\cdots$) elements,
it should, however, not be a surprise that  $\bar{S}_{\text{q}}$ (Fig.~\ref{fig_synchronization}) and $\bar{N}$ (Fig.~\ref{figure_adaggera})
display similar overall characteristics.

To quantify the deformation of the limit cycle due to the external drive, Fig.~\ref{figure_dbar} shows the 
time-averaged deformation $\bar{D}$ for the same parameters as those employed in Figs.~\ref{fig_synchronization}, \ref{figure_adaggera}, and \ref{figure_dbar}. 
Not surprisingly, the overall behavior of $\bar{D}$ resembles---just as that of $\bar{S}_{\text{q}}$ and $\bar{N}$---an Arnold tongue. While the deformation is quite large for ``large" drive strengths and ``small" detunings, inspection of the Wigner functions shows that the limit cycle is not broken, i.e., the maximum of the Wigner function still follows the shape of the zero-drive limit cycle, though with increased amplitude.
 Even though the deformation $\bar{D}$ is appreciable, we are operating in a regime where the phase localization is intimately connected to the zero-drive limit cycle.  

Next, we discuss the spectral response.
The spectra shown in Figs.~\ref{figure_spectra} and \ref{figure_spectra2}  are calculated within the RWA in the frame that rotates at the drive frequency. 
We use Eq.~(\ref{eq_powerspectrum}) with $\tau=200$ and simulation parameters that yield $\omega \in [-3,3]$ with a frequency resolution of 
$1/500$. 
Consistent with the $\Omega=0$ limit cycle discussion presented earlier, $S_p(\omega,\tau)$ is independent of $\tau$ for the RvdP
oscillator (middle column of Figs.~\ref{figure_spectra} and \ref{figure_spectra2}) 
but displays, in general, a dependence on $\tau$ for the R and vdP oscillators (left and right columns of Figs.~\ref{figure_spectra} and \ref{figure_spectra2}).
The
$\tau$-dependence tends to be weak (similar behavior was observed in Ref.~\cite{chia2020}; see also discussion below).

The top row of Fig.~\ref{figure_spectra} shows the power spectrum for the R, RvdP, and vdP oscillators in the quantum regime as functions of $\omega$ and $\Delta$  [the linear and non-linear parameters are the same as those employed in
Figs.~\ref{fig_synchronization}(k), \ref{fig_synchronization}(m), and \ref{fig_synchronization}(o)].  
We note that beyond the RWA terms, which are not included in the calculations, might play a non-negligible role for the larger $|\Delta|$.
The spectra 
are, for each $\Delta$, normalized such that the maximum of the broad peak that is centered at $\omega \approx \Delta$ is equal to $1$. For all $\Delta$ considered, the height of the $\delta$-function-like peak is larger than one, i.e., the value of this sharp spike is capped for visualization purposes.
 The red (lowest) curves in Figs.~\ref{figure_spectra}(d)-\ref{figure_spectra}(f)
  correspond to horizontal $\Delta=0$ cuts through the data shown in Figs.~\ref{fig_synchronization}(k), \ref{fig_synchronization}(m), and \ref{fig_synchronization}(o). For comparison, the green (middle) curves and blue (top) curves show $\Delta=0$ spectra for the intermediate and classical regimes, respectively, using the same non-linear parameters as in Figs.~\ref{fig_synchronization}(f), ~\ref{fig_synchronization}(h), ~\ref{fig_synchronization}(j), ~\ref{fig_synchronization}(a), ~\ref{fig_synchronization}(c), and ~\ref{fig_synchronization}(e). Unlike in the top row,  the maximum of the broad peak is not scaled to $1$ in the bottom row of Fig.~\ref{figure_spectra}. Moreover, the $\delta$-function-like spike 
  (a single data point at $\omega=0$) is taken out by hand.
  For comparison,
  Fig.~\ref{figure_spectra2} shows the power spectra $S_p(\omega,\tau)$ for a finite detuning, namely $\Delta=1/20$, and two different drive strengths ($\Omega=3\sqrt{2}/5 \approx 0.849$ and $\sqrt{2}/5 \approx 0.283$ for the top 
  and bottom rows, respectively) using the same linear and non-linear parameters as well as the same color coding as Fig.~\ref{figure_spectra}. The insets of Fig.~\ref{figure_spectra2} 
  omit the $\delta$-function-like spike and normalize, as in the top row of Fig.~\ref{figure_spectra}, the spectra such that the broad peak has a height of one. The insets allow one to read off at which $\omega$ the maximum of the broad peak is located.

The key characteristics of the power spectra are:
(i) The very sharp peak centered at $\omega=0$ (top row of Fig.~\ref{figure_spectra} and main panels of Fig.~\ref{figure_spectra2})
exists for all three oscillator types; since we are working in the frame that rotates with the drive frequency, this peak reflects a strong response at the frequency corresponding to the drive. 
(ii) The broad peak centered at $\omega \approx \Delta$ (see, e.g., the bright red feature in the top row of Fig.~\ref{figure_spectra}) exists for all three oscillator types; this broad peak can be interpreted as a strong response, broadened by the dissipative processes, around the natural harmonic oscillator frequency. 
As discussed in more detail below,
the maximum of this broad peak is, in general, not located at $\omega = 0$; thus, true entrainment is, in general, absent.
(iii) The broad peak is narrower for the RvdP oscillator than for the R and vdP oscillators. Moreover, the broadness of the peak depends rather weakly on $\Delta$. The most pronounced dependence on $\Delta$ can be seen for the vdP oscillator, for which the power spectrum depends most strongly on $\tau$. A period average of a sequence of power spectra for different $\tau$ would smooth out the horizontal  stripes in Fig.~\ref{figure_spectra}(c).
(iv) The broad peak centered at $\omega \approx \Delta$ is broader in the quantum regime than in the classical regime for all oscillator types.
This is attributed to the increase of the zero-point motion in the quantum regime~\cite{arosh2021,walter2014}. 
(v) The R and vdP oscillators feature sharp peaks, corresponding to higher harmonics, that are centered at
$\omega=2 \Delta+2$, 
$\omega=2\Delta-4$,
and $\omega=2\Delta-2$ 
[in the quantum regime, this peak is hardly visible
on the scale shown in Figs.~\ref{figure_spectra}(d) and \ref{figure_spectra}(f)].

\begin{figure}[t]
\includegraphics[width=0.42\textwidth]{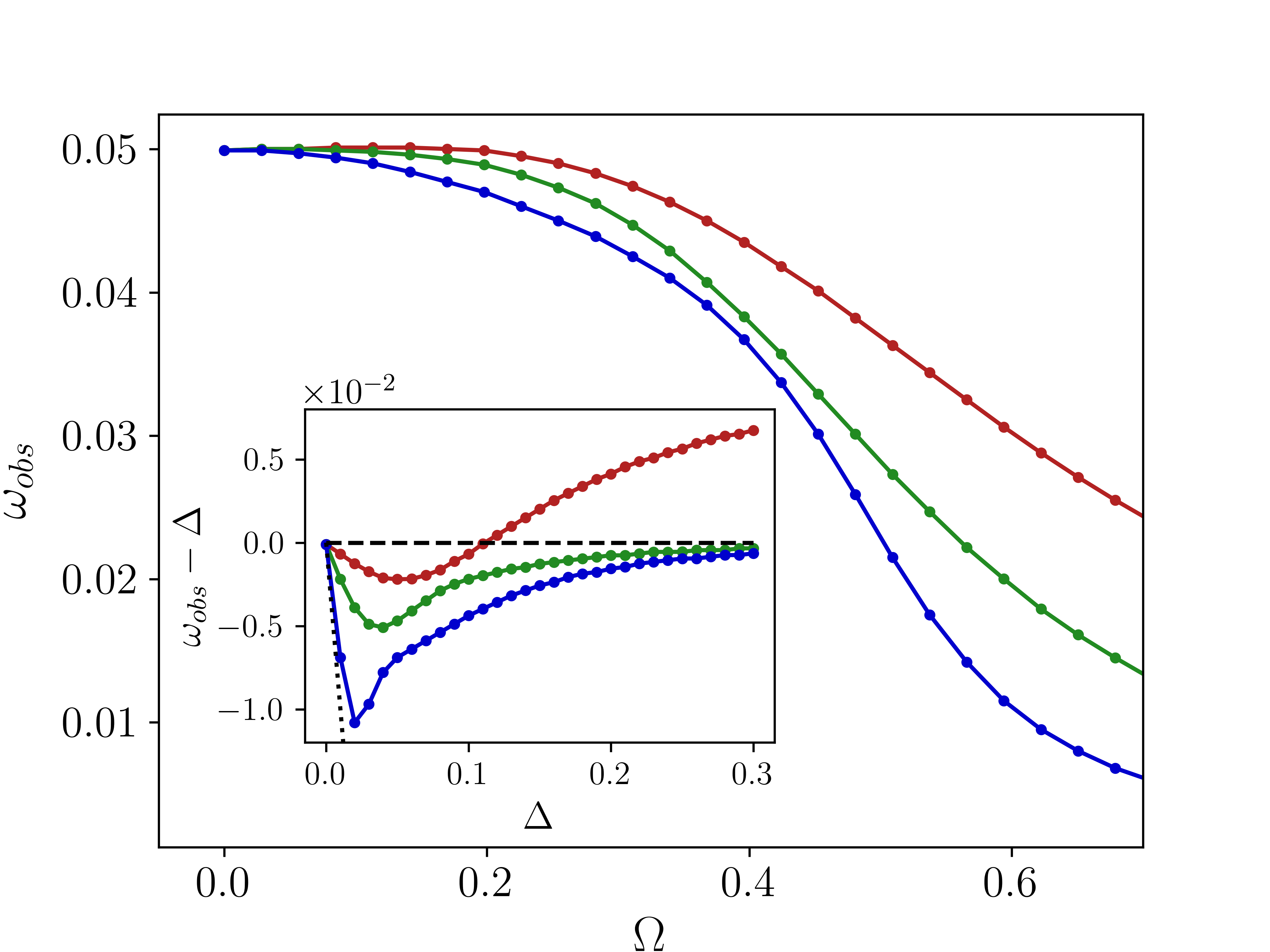}
\caption{Analysis of the maximum of the broad peak for the RvdP oscillator for $\tau=200$.
Main panel: Blue (bottom), green (middle), and red (top) curves show the
location $\omega_{\text{obs}}$ of the maximum of the broad peak 
for the classical regime, transition regime, and quantum regime, respectively, as a function of the drive strength $\Omega$ for $\Delta=1/20$.
The linear parameters are $\gamma_1^+=1/5$ and $\gamma_1^-=1/10$ while 
the non-linear parameters for the red (bottom) curve, green (middle) curve, and  blue (top) curve are given in 
 Table~\ref{table_parameters_fig1}(m), \ref{table_parameters_fig1}(h),
 and \ref{table_parameters_fig1}(c), respectively. 
The  calculations are performed within the RWA in the frame that rotates at the drive frequency.
Inset: 
Blue (bottom), green (middle), and red (top) curves show the
shifted location $\omega_{\text{obs}}-\Delta$ of the maximum of the broad peak 
for the classical regime, transition regime, and quantum regime, respectively, as a function of the detuning $\Delta$ for $\Omega=3/10$ (the other parameters are the same as those in the main panel). If $\omega_{\text{obs}}-\Delta$ followed the black dotted line, the system would exhibit true entrainment. On the other hand, if $\omega_{\text{obs}}-\Delta$ followed the black dashed line, the system response would be governed by the natural harmonic oscillator frequency. }
\label{figure_spectra3}
\end{figure}

To gain insights into the existence or lack of entrainment for small $|\Delta|$, 
we investigate the difference between the frequency
$\omega_{\text{obs}}$, at which the broad peak that is centered around
$\omega \approx \Delta$
takes its maximum, and  the detuning $\Delta$.
This difference can be read off the spectra, such as those shown 
in Figs.~\ref{figure_spectra}(d)-\ref{figure_spectra}(f) for zero detuning and
the insets of Fig.~\ref{figure_spectra2} for finite detuning.
The main panel of   
Fig.~\ref{figure_spectra3} shows $\omega_{\text{obs}}$ as a function of the drive strength
$\Omega$ for the RvdP oscillator in the quantum, transition, and classical regimes
[curves from bottom (blue) to top (red)] for the same linear and non-linear parameters as
used in the middle columns of Figs.~\ref{figure_spectra} and \ref{figure_spectra2}.
For all three regimes, $\omega_{\text{obs}}$ is equal to $\Delta$ for $\Omega=0$
(in the absence of the drive, the spectral response is maximal at the natural harmonic
oscillator frequency) and decreases monotonically with increasing $\Omega$.
Even though $\omega_{\text{obs}}$ tends toward zero, especially in the classical regime
(a value of $\omega_{\text{obs}}=0$ would indicate entrainment), true entrainment is absent for the oscillator strengths considered. The drive strength was not increased further since we wish to remain in the
weakly-perturbed regime, where the driven system inherits key characteristics of the 
limit-cycle state of the undriven system.

To further explore the presence or absence of entrainment, the inset of Fig.~\ref{figure_spectra3}
shows $\omega_{\text{obs}}-\Delta$ as a function of the detuning for the RvdP oscillator.
Using $\omega_{\text{obs}}$,  as done in Ref.~\cite{walter2014}, as a proxy for the frequency at which the system responds 
to, entrainment would correspond to $\omega_{\text{obs}}-\Delta$ following the dashed line. It can be seen that 
entrainment exists, for sufficiently small $\Delta$, in the classical regime but not in the quantum regime (of course, we cannot rule out the existence of entrainment in this regime for $|\Delta|$ values that are below our resolution scale).
The observation of entrainment in the classical regime is consistent with
results presented in  Ref.~\cite{walter2014}.
The  ``oscillatory" behavior of $\omega_{\text{obs}}-\Delta$ in the quantum regime is interpreted as reflecting an appreciable ``hybridization" of the spectral function with regards 
to the drive and oscillator frequencies. This deep quantum regime has not,
to the best of our knowledge, been investigated previously.
We finish this section with the following comments.
In the limit where the broad peak is comparatively narrow, i.e., where the peak width is smaller or of the same order of magnitude as $|\Delta|$, $\omega_{\text{obs}}$ is a reliable indicator of entrainment.
For broad system responses, such as those observed in the deep quantum regime, however, it might 
be more meaningful 
to focus on the entire spectral response as opposed to a single feature thereof.

\section{Conclusions}
\label{sec_summary}
This paper presented a detailed analysis of the driven generalized Rayleigh-van der Pol 
oscillators---including the driven Rayleigh (R), driven Rayleigh-van der Pol (RvdP),
and driven van der Pol (vdP) oscillators---within a master equation framework. 
While the drive was treated classically, the motional degrees of freedom of the oscillator were
quantized, with linear and non-linear loss/gain accounted for through dissipators.
In the weak $\epsilon$-limit (i.e., for small effective linear loss or gain), a mapping to the classical
equations of motion was established. The mapping allows for a transparent interpretation of the
dissipative terms. An important aspect of the generalized RvdP oscillator is that the dissipators
are not, in general, rotationally symmetric in phase space. As a consequence, the drive-free 
Wigner functions exhibit a phase-space asymmetry that increases with
decreasing number $\bar{N}$ of excitations. Our main focus throughout was to characterize the 
quasi-stationary long-time regime, where the dynamics is linked to the limit-cycle of 
the undriven system; this limit cycle emerges due to the competition between
linear and non-linear dissipative processes.

In classical settings,  phase localization and frequency 
entrainment are critical elements for harnessing characteristics associated with
limit-cycles; correspondingly, classical synchronization is defined as requiring
phase localization and frequency 
entrainment.
By analogy, we defined quantum synchronization as requiring
phase localization and frequency 
entrainment.
Our analysis shows that the driven generalized RvdP oscillators 
exhibit Arnold-like synchronization tongues. 
This work employed a phase operator-based definition for
the phase localization measure $S_{\text{q}}$.
Observables such as the excitation number and the deformation
of the limit cycle were found to exhibit analogous Arnold-tongue shaped characteristics, if plotted
as functions of the detuning and drive strength. 
Phase localization was found to decrease for all oscillator types
considered as the system changed from the
classical to the quantum regime.
The phase space  asymmetry of the non-RvdP oscillators was found to reduce 
phase localization.
In the quantum regime, where the 
number of excitations $\bar{N}$ is low, the spectral response of the systems was found to be 
extremely broad, suggesting that the frequency at which the non-delta function
response of the spectrum is maximal provides only limited information about the system's response.
While appreciable phase space localization was observed  in the deep quantum regime, 
weak frequency entrainment was only found for the RvdP oscillator in the deep quantum regime only 
for a narrow range of
small, in absolute value, detunings. 
We were not able to identify a parameter regime in the deep quantum regime
where the driven generalized RvdP oscillator exhibits true non-trivial synchronization, 
i.e., phase  localization and frequency locking to the drive.

{\em{Acknowledgement:}}
We gratefully
acknowledge fruitful discussions with Xylo Molenda as well as the Biedermann and Marino groups.
This work was supported by an award from the W.~M. Keck Foundation.
Partial support by the National Science Foundation through grant no. PHY-1950235 (REU/RET) and OU's H-RAP program is acknowledged. 
This paper was completed during the KITP program 
“Out-of-equilibrium Dynamics and Quantum Information of Many-body Systems with Long-range Interactions;" partial support by grants NSF PHY-1748958 and PHY-2309135 to the Kavli Institute for Theoretical Physics (KITP) is acknowledged. 
This work used the OU Supercomputing Center for Education and Research (OSCER) at the University of Oklahoma (OU).

\appendix

\section{Master equation in the rotating frame}
\label{appendix_rotating}

To get started, we
introduce the rotation operator $\hat{U}(t)$,
\begin{eqnarray}
\hat{U}(t)=e^{-\imath \omega_R \hat{a}^{\dagger} \hat{a} t},
\end{eqnarray}
where the angular frequency $\omega_R$ could be equal to the angular frequency of the harmonic oscillator
(which is equal to $1$ in the dimensionless units employed throughout), the angular frequency $\omega_D$ of the
drive, or some other angular frequency. 
The operator $\hat{C}$ in the rotating frame   is denoted by $\hat{C}_R$,
\begin{eqnarray}
\hat{C}_R=
\hat{U}^{\dagger}(t) \hat{C} \hat{U}(t).
\end{eqnarray}
For the driving term, one finds
\begin{eqnarray}
(\hat{V}_{\text{drive}})_R = 
\frac{  \Omega}{2\imath  \sqrt{2}}
\left(
e^{\imath (\omega_D-\omega_R)t } \hat{a}
-
e^{-\imath (\omega_D-\omega_R)t } \hat{a}^{\dagger}
\right) 
+
\nonumber \\
\frac{  \Omega}{2 \imath \sqrt{2}}
\left(
e^{\imath (\omega_D+\omega_R)t } \hat{a}^{\dagger}
-
e^{-\imath (\omega_D+\omega_R)t } \hat{a}
\right),
\end{eqnarray}
where the terms on the right hand side correspond to terms within the RWA (first line) and terms beyond the RWA (second line), respectively.

The master equation in the frame rotating with angular frequency $\omega_R$ reads
\begin{eqnarray}
\dot{\hat{\rho}}_R = 
-\imath [(1-\omega_R)\hat{H}_0+(\hat{V}_{\text{drive}})_R,\hat{\rho}_R]
+ \nonumber \\
\gamma_1^+ \hat{{\cal{D}}} [\hat{a}_R^{\dagger} ] (\hat{\rho}_R)
+ 
\gamma_1^- \hat{{\cal{D}}} [\hat{a}_R ] (\hat{\rho}_R)
+
\alpha \hat{{\cal{D}}} [\hat{a}_R \hat{a}_R ] (\hat{\rho}_R)
+ \nonumber \\
\beta \hat{{\cal{D}}} [\hat{x}_R \hat{a}_R ] (\hat{\rho}_R)
+ 
\delta \hat{{\cal{D}}} [\hat{p}_R \hat{a}_R ] (\hat{\rho}_R),
\end{eqnarray}
where 
\begin{eqnarray}
\hat{{\cal{D}}} [\hat{C}_R](\hat{\rho}_R) = \hat{U}^{\dagger}(t)
\hat{{\cal{D}}} [ \hat{C}](\hat{\rho})
\hat{U}(t).
\end{eqnarray}
It is readily shown that the form of the dissipators that are proportional to $\gamma_1^+$, $\gamma_1^-$, and $\gamma_2$ is the same in the rotating frame as in the laboratory frame:
\begin{eqnarray}
\hat{{\cal{D}}} [ \hat{a}_R] (\hat{\rho}_R)
=
\hat{{\cal{D}}} [\hat{a}] (\hat{\rho}_R),
\end{eqnarray}
\begin{eqnarray}
\hat{{\cal{D}}} [ \hat{a}_R^{\dagger}] (\hat{\rho}_R)
=
\hat{{\cal{D}}} [\hat{a}^{\dagger}] (\hat{\rho}_R),
\end{eqnarray}
and
\begin{eqnarray}
\hat{{\cal{D}}} [ \hat{a}_R \hat{a}_R ] (\hat{\rho}_R)
=
\hat{{\cal{D}}} [\hat{a} \hat{a}] (\hat{\rho}_R).
\end{eqnarray}
By ``same form" (``different form") we mean that the argument of $\hat{\cal{D}}$ takes the same form (a different form) in the rotating frame as that in the laboratory frame.
The dissipators that are proportional to $\beta$ and $\delta$, in contrast, take a different form in the rotating frame than in the laboratory frame: 
\begin{eqnarray}
\hat{{\cal{D}}} [\hat{x}_R \hat{a}_R] (\hat{\rho}_R)
=
\hat{{\cal{D}}} [(\hat{x}+\hat{y}+\hat{y}^{\dagger})\hat{a}] (\hat{\rho}_R)
\end{eqnarray}
and
\begin{eqnarray}
\hat{{\cal{D}}} [ \hat{p}_R \hat{a}_R] (\hat{\rho}_R)
=
\hat{{\cal{D}}} [\left(\hat{p} - \imath (\hat{y}-\hat{y}^{\dagger})\right)\hat{a}] (\hat{\rho}_R),
\end{eqnarray}
where the time-dependent operator $\hat{y}$ is defined as
\begin{eqnarray}
\hat{y}(t) = \frac{1}{\sqrt{2}} \left[
\exp(-  \imath \omega_R t) -1
\right] \hat{a}.
\end{eqnarray}
The appearance of the operators $\hat{y}$ and 
$\hat{y}^{\dagger}$ may be interpreted as being due to a fictitious force in the rotating frame.
For in-depth discussions of the connection between the laboratory- and rotating-frame master equations, the reader is referred to Refs.~\cite{shavit2019,munro1996,baker2018}.

\section{Phase localization measures}
\label{appendix_measure}
Several quantum phase localization measures have been proposed (see, e.g., Refs.~\cite{tindall2020,thomas2022,ameri2015,jaseem2020,mari2013,lee2014,galve2017,barak2005,heimonen2018,zhirov2006}), including measures based on the phase operator and quantum information theory frameworks, not only in the context of different oscillator types---as considered in this work---but also in the context of spin systems~\cite{goychuk2006,zhirov2008,giorgi2013,xu2014,hush2015,ameri2015,roulet2018,roulet2018a,koppenhofer2019,tan2022}. This appendix provides more context for the phase localization measure $S_q$, Eq.~(\ref{eq_synchronization_raw}), employed in this work and relates it to other phase operator based measures.

The phase operator $\hat{\varphi}$ is defined through $\hat{\varphi}|\varphi \rangle = \varphi | \varphi \rangle$, where
the phase state $|\varphi\rangle$ is a superposition of the harmonic oscillator eigen states 
$|n\rangle$~\cite{pegg1988}, 
\begin{eqnarray}
|\varphi \rangle = \frac{1}{\sqrt{2 \pi}}
\sum_{n=0}^{\infty} \exp ( \imath n \varphi )| n \rangle.
\end{eqnarray}
Using the properties of the phase states,
Eq.~(\ref{eq_synchronization}) follows readily from Eq.~(\ref{eq_synchronization_raw}).

An alternative phase localization measure reads
$S_{\text{q},\text{alter1}}=\Re(\langle \hat{a}\rangle)/|\langle \hat{a} \rangle|$~\cite{weiss2016}, where $\Re(C)$ denotes the real part of $C$. Writing $\langle \hat{a}\rangle$ in terms of the amplitude $|a|$ and phase $\varphi$ [namely, $\langle \hat{a}\rangle=|a|\exp(-\imath \varphi)$] highlights that $\langle \hat{a}\rangle$ contains phase information. 
Evaluating $S_{\text{q},\text{alter1}}$ in the harmonic oscillator basis, 
one finds
\begin{eqnarray}
\label{eq_alter1}
S_{\text{q},\text{alter1}}=\frac{\Re \left(
\sum_{n=1}^{\infty} \sqrt{n} \rho_{n,n-1} \right)}
{\left| \sum_{n=1}^{\infty} \sqrt{n} \rho_{n,n-1} \right|}.
\end{eqnarray}
Equation~(\ref{eq_alter1}) shows 
that the density matrix elements that contribute to $S_{\text{q},\text{alter1}}$ are the same as those that contribute to
$S_{\text{q}}$; the density matrix elements are, however, weighted differently. Use of the imaginary part of $\langle \hat{a} \rangle$ yields complementary insights~\cite{weiss2016}.

We note that $S_{\text{q},\text{alter1}}$ can, in the small $\Omega$ limit, be related to the susceptibility $\chi$~\cite{dutta2019},
\begin{eqnarray}
\chi = \frac{\partial \langle \hat{a} \rangle}{\partial \Omega}.
\end{eqnarray}
To see the connection between $S_{\text{q},\text{alter1}}$ and $\chi$, we employ the perturbation theory
framework detailed in Ref.~\cite{koppenhofer2019}.
When the $\Omega=0$ steady state density matrix is used as zeroth-order solution and the external drive in the RWA is treated as a perturbation,  $\langle \hat{a} \rangle $ is directly proportional to $\Omega$.  
It follows that $S_{\text{q},\text{alter1}}$ is 
equal to $\Re(\Omega \chi)/|\langle \hat{a} \rangle |$ in the weakly-driven regime, where first-order perturbation theory provides a faithful description.
This indicates that the susceptibility results  for the driven 
RvdP 
oscillator for small $\Omega$~\cite{dutta2019}  can be interpreted within the framework of phase localization. 
Note, however, that since phase localization requires the presence of a non-linearity, only the results
from Ref.~\cite{dutta2019} for
non-vanishing non-linearity can be meaningfully reinterpreted in this way.

Phase localization can also be defined in terms of the normalized phase probability $P(\varphi)$~\cite{hush2015},
\begin{eqnarray}
P(\varphi)=  \int_0^{2 \pi} \delta(\varphi-\varphi') 
\langle \varphi' | \hat{\rho}| \varphi'\rangle d \varphi' .
\end{eqnarray}
Using the maximum of the normalized phase
probability, the phase localization measure
$S_{\text{q},\text{alter2}} $, which is frequently used in the context of spin systems, reads~\cite{hush2015}
\begin{eqnarray}
S_{\text{q},\text{alter2}} = 2 \pi \max_{\varphi}P(\varphi) - 1.
\end{eqnarray}
Since $\langle \hat{\rho} \rangle$ is equal to $1$,
$S_{\text{q},\text{alter2}} $ reduces to
\begin{eqnarray}
S_{\text{q},\text{alter2}}=
\max_{\varphi}
 \sum_{k \ne l} \exp[ -\imath (k-l)\varphi]
\rho_{k,l}.
\end{eqnarray}
Using that $\rho_{k,l}$ is equal to
$(\rho_{l,k})^*$, it can be shown readily that $S_{\text{q},\text{alter2}}$ is real.
Physically, $S_{\text{q},\text{alter2}}$ can be thought of as extracting the maximal height of $P(\varphi)$.
Note that $S_{\text{q},\text{alter2}}$ is sensitive to all off-diagonal density matrix elements. Specifically, the  operation $\max_{\varphi}$ ``optimizes" the multiplicative factor of the sum of the density matrix elements with $|k-l|=1$ [multiplicative factor is $\exp(\imath \varphi)$], $|k-l|=2$ [multiplicative factor is $\exp(2 \imath  \varphi)$], and so on.

We now discuss the $\beta=\delta$ oscillator, treating  the drive  in the RWA [see
Eq.~(\ref{eq_drive_RWA})], in more detail. For the arguments that follow, it is convenient to work in the frame that is rotating with the external drive. For $\beta=\delta$,  the $\Omega=0$ steady state density matrix $\rho^{(0)}$ is---as already mentioned in Sec.~\ref{sec_theory}---diagonal.
It is determined by 
$\hat{{\cal{L}}}_0 \hat{\rho}^{(0)}=0$, where $\hat{{\cal{L}}}_0$ denotes the  superoperator that corresponds to the $\Omega=0$ master equation~\cite{koppenhofer2019}. 
The first-order density matrix $\hat{\rho}^{(1)}$
is obtained by solving 
$\hat{{\cal{L}}}_{\text{drive}} \hat{\rho}^{(0)}=-\hat{{\cal{L}}}_0 \hat{\rho}^{(1)}$,
where the superoperator $\hat{\cal{L}}_{\text{drive}}$
is fully determined by the external drive~\cite{koppenhofer2019}.  
It can be readily shown that the only non-zero elements of $\hat{\rho}^{(1)}$ are those with $|k-l|=1$. Consequently,
$S_{\text{q},\text{alter2}}$ reduces to
$2 \max_{\varphi} \left[ \Re(\cos \varphi \sum_{n=1}^{\infty}\rho_{n,n-1})
+ \Im(\sin \varphi \sum_{n=1}^{\infty}\rho_{n,n-1}) \right]$.
This shows that  $S_{\text{q}}$,  $S_{\text{q},\text{alter1}}$, and $S_{\text{q},\text{alter2}}$ are all governed by the same
matrix elements.

Releasing the restriction that $\Omega$ is small while continuing to focus on the 
RvdP
oscillator 
($\beta=\delta$), another interesting limit is the deep quantum regime
(characterized by small $\langle \hat{a}^{\dagger} \hat{a} \rangle$), 
where the full steady state density matrix elements  are well approximated by 
a three-state model~\cite{mok2020}. Within the three-state model, the only non-zero off-diagonal elements are $\rho_{1,0}=(\rho_{0,1})^*$. Correspondingly,
$S_{\text{q}}$, $S_{\text{q},\text{alter1}}$, and $S_{\text{q},\text{alter2}}$ are related to each other in a straightforward manner.


\begin{thebibliography}{100}

\bibitem{pikovsky_book}
A. Pikovsky, M. Rosenblum, and J. Kurths, “Synchronization. A Universal Concept in Nonlinear Sciences," Cambridge, Cambridge University Press (2001).

\bibitem{balanov_book}
A. Balanov, N. Janson, D. Postnov, and O. Sosnovtseva,
“Synchronization,"
Springer, Berlin/Heidelberg, 2009.

\bibitem{strutt1883}
J. W. Strutt and Baron Rayleigh, “On maintained vibrations,” Philos. Mag. (ser. 5) {\bf{15}}, 229-235 (1883)

 \bibitem{vanderpol1}
B. van der Pol, “Forced oscillators in a circuit with non-linear resistance. (Reception with reactive triode)," Phil. Mag. {\bf{3}}, 64 (1927).

\bibitem{vanderpol2}
 B. van der Pol and J. van der Mark, “The Heartbeat considered as a Relaxation oscillation, and an Electrical Model of the Heart," Phil. Mag. Suppl. {\bf{6}},  763 (1928). 
 

\bibitem{footnote_conversion}
We start with the differential equation for the Rayleigh oscillator [Eq.~(\ref{eq_classical_eom}) with $\Omega=\bar{\gamma}_{2,\text{vdp}}=0$].
Differentiation with respect to time yields:
$\dddot{x}+\dot{x}
= \epsilon ( 1- 4 \bar{\gamma}_{2,\text{ray}} \dot{x}^2)\ddot{x}$.
Defining $\bar{\gamma}_{2,\text{vdp}}=4 \bar{\gamma}_{2,\text{ray}}$ and making the substitution $y=\dot{x}$, one obtains the differential equation for the van der Pol oscillator.


\bibitem{chia2020}
A. Chia, L. C. Kwek, and C. Noh,
“Relaxation oscillations and frequency entrainment in quantum mechanics,"
Phys. Rev. E {\bf{102}}, 042213 (2020).

\bibitem{arosh2021}
L. B. Arosh, M. C. Cross, and R. Lifshitz,
“Quantum limit cycles and the Rayleigh and van der Pol oscillators,"
Phys. Rev. Research {\bf{3}}, 013130 (2021).

\bibitem{lee2013}
T. E. Lee and H. R. Sadeghpour,
“Quantum Synchronization of Quantum van der Pol Oscillators with Trapped Ions,"
Phys. Rev. Lett. {\bf{111}}, 234101
 (2013).

\bibitem{walter2014}
S. Walter, A. Nunnenkamp, and C. Bruder,
“Quantum Synchronization of a Driven Self-Sustained Oscillator,"
Phys. Rev. Lett. {\bf{112}}, 094102 (2014).



\bibitem{lorch2017}
N. L\"orch, S. E. Nigg, A. Nunnenkamp, R. P. Tiwari, and C. Bruder,
“Quantum Synchronization Blockade: Energy Quantization Hinders Synchronization of Identical Oscillators,"
Phys. Rev. Lett. {\bf{118}}, 243602 (2017).

\bibitem{morgan2015}
L. Morgan and H. Hinrichsen,
“Oscillation and synchronization of two quantum self-sustained oscillators,"
J. Stat. Mech.: Theory Exp. 09, P09009 (2015).

\bibitem{lee2014}
T. E. Lee, C.-K. Chan, and S. Wang,
“Entanglement tongue and quantum synchronization of disordered oscillators,"
Phys. Rev. E {\bf{89}}, 022913 (2014).

\bibitem{amitai2017}
E. Amitai, N. L\"orch, A. Nunnenkamp, S. Walter, and C. Bruder,
“Synchronization of an optomechanical system to an external drive,"
Phys. Rev. A {\bf{95}}, 053858 (2017).

\bibitem{amitai2018}
E. Amitai, M. Koppenh\"ofer, N. L\"orch,  and C. Bruder,
“Quantum effects in amplitude death of coupled anharmonic self-oscillators,"
Phys. Rev. E {\bf{97}}, 052203 (2018).


\bibitem{hush2015}
M. R. Hush, W. Li, S. Genway, I. Lesanovsky, and A. D. Armour,
“Spin correlations as a probe of quantum synchronization in trapped-ion phonon lasers,"
Phys. Rev. A {\bf{91}} 061401(R) (2015).

\bibitem{ameri2015}
V. Ameri, M. Eghbali-Arani, A. Mari, A. Farace, F. Kheirandish, V. Giovannetti, and R. Fazio,
“Mutual information as an order parameter for quantum synchronization,"
Phys. Rev. A {\bf{91}}, 012301 (2015).



\bibitem{weiss2017}
T. Weiss, S. Walter, and F. Marquardt,
“Quantum-coherent phase oscillations in synchronization,"
Phys. Rev. A {\bf{95}}, 041802(R) (2017).

\bibitem{sonar2018}
S. Sonar, M. Hajdu{\u{s}}ek, M. Mukherjee, R. Fazio, V. Vedral, S. Vinjanampathy, and L.-C. Kwek,
“Squeezing Enhances Quantum Synchronization,"
Phys. Rev. Lett. {\bf{120}} 163601 (2018).

\bibitem{koppenhofer2019}
M. Koppenh\"ofer and A. Roulet,
“Optimal synchronization deep in the quantum regime: Resource and fundamental limit,"
Phys. Rev. A {\bf{99}}, 043804 (2019).

\bibitem{mok2020}
W.-K. Mok, L.-C. Kwek, and H. Heimonen,
“Synchronization boost with single-photon dissipation in the deep quantum regime,"
Phys. Rev. Research {\bf{2}}, 033422 (2020).



\bibitem{jaseem2020}
N. Jaseem, and M. Hajdu{\u{s}}ek, P. Solanki, L.-C. Kwek, R. Fazio, and S. Vinjanampathy,
“Generalized measure of quantum synchronization,"
Phys. Rev. Research {\bf{2}}, 043287 (2020).

\bibitem{li2020}
J. Li, C. Ding, and Y. Wu,
“Highly nonclassical phonon emission statistics through two-photon loss of van der Pol oscillator,"
J. Appl. Phys. {\bf{128}}, 234302 (2020).

\bibitem{cabot2021}
A. Cabot, G. C. Giorgi, and R. Zambrini,
“Metastable quantum entrainment,"
New J. Phys. {\bf{23}}, 103017 (2021).

\bibitem{thomas2022}
N. Thomas and M. Senthilvelan,
“Quantum synchronization in quadratically coupled quantum van der Pol oscillators,"
Phys. Rev. A {\bf{106}}, 012422 (2022).

\bibitem{eshaqi2020}
N. Es'haqi-Sani, G. Manzano, R. Zambrini, and R. Fazio,
“Synchronization along quantum trajectories,"
Phys. Rev. Research {\bf{2}}, 023101 (2020).

\bibitem{dutta2019}
S. Dutta and N. R. Cooper,
“Critical Response of a Quantum van der Pol Oscillator,"
Phys. Rev. Lett. {\bf{123}}, 250401 (2019).

\bibitem{simaan1975}
H. D. Simaan and R. Loudon,
“Quantum statistics of single-beam two-photon absorption,"
J. Phys. A: Math. Gen. {\bf{8}}, 539 (1975).

\bibitem{dodonov1997}
V. V. Dodonov and S. S. Mizrahi,
“Exact stationary photon distributions due to competition between one- and two-photon absorption and emission,"
J. Phys. A: Math. Gen. {\bf{30}}, 5657 (1997).

\bibitem{footnote1}
The master equation for $\beta=\delta \ne 0$ is equivalent to that for $\beta=\delta=0$ and modified $\alpha$: $\alpha \rightarrow \alpha+\beta/2+\delta/2=\alpha+\beta$.

\bibitem{quantum_optics_book}
M. O. Scully and M. S. Zubairy,
Quantum Optics, 1st Edition, Cambridge University Press, 1997.

\bibitem{bender_book}
C. M. Bender and S. A. Orszag,
“Advanced Mathematical Methods for Scientists and Engineers: Asymptotic Methods and Perturbation Theory,"
Springer, 1999th Edition.

\bibitem{burton1984}
T. D. Burton,
“A perturbation method for certain non-linear oscillators,"
Int. J. Non-Linear Mechanics {\bf{19}}, 397 (1984).

\bibitem{goldstein_book}
H. Goldstein, C. Poole, and J. Safko,
“Classical Mechanics,"
3rd Edition, Addison Wesley.

\bibitem{mari2013}
A. Mari, A. Farace, N. Didier, V. Giovannetti, and R. Fazio,
“Measures of Quantum Synchronization In Continuous Variable Systems,"
Phys. Rev. Lett. {\bf{111}}, 103605 (2013).

\bibitem{tindall2020}
J. Tindall, C. S\'anchez Mu{\~n}oz, B. Bu{\u{c}}a, and D. Jaksch,
“Quantum synchronization enabled by dynamical symmetries and dissipation,"
New J. Phys. {\bf{22}}, 013026 (2020).

\bibitem{pegg1988}
D. T. Pegg and S. M. Barnett,
“Unitary Phase Operator in Quantum Mechanics," 
EPL {\bf{6}}, 483 (1988).

\bibitem{shapiro1991}
J. H. Shapiro and S. R. Shepard,
“Quantum phase measurement: A system-theory perspective," Phys. Rev. A {\bf{43}}, 3795 (1991).

\bibitem{breuer2002}
H. Breuer and F. Petruccione, 
“The Theory of Open Quantum Systems," Oxford University Press, New York, 2002.

\bibitem{weiss2016}
T. Weiss, A. Kronwald, and F. Marquardt,“Noise-induced transitions in optomechanical synchronization,"
New J. Phys. {\bf{18}}, 013043 (2016).

\bibitem{andre2012}
S. Andr\'e, L. Guo, V. Peano, M. Mathaler, and G. Sch\"on,
“Emission spectrum of the driven nonlinear oscillator,"
Phys. Rev. A {\bf{85}}, 053825 (2012).

\bibitem{lee2013a}
T. E. Lee and M. C. Gross,
“Quantum-classical transition of correlations of two coupled cavities,"
Phys. Rev. A {\bf{88}}, 013834 (2013).

\bibitem{shavit2019}
G. Shavit, B. Horowitz, and M. Goldstein,
“Bridging between laboratory and rotating-frame master equations for open quantum systems,"
Phys. Rev. B {\bf{100}}, 195436 (2019).

\bibitem{munro1996}
W. J. Munro and C. W. Gardiner,
“Non-rotating-wave master equation,"
Phys. Rev. A {\bf{53}}, 2633 (1996).

\bibitem{baker2018}
B. Baker, A. C. Y. Li, N. Irons, N. Earnest, and J. Koch,
“Adaptive rotating-wave approximation for driven open quantum systems,"
Phys. Rev. A {\bf{98}}, 052111 (2018).

\bibitem{galve2017}
F. Galve, G.-L. Giorgi, and R. Zambrini, “Quantum correlations and synchronization measures," in Lectures on General Quantum Correlations and Their Applications, edited by F. F. Fanchini, D. d. O. S. Pinto, and G. Adesso (Springer, Berlin, 2017), pp. 393–420.

\bibitem{heimonen2018}
H. Heimonen, L. C. Kwek, R. Kaiser, and G. Labeyrie,
“Synchronization of a self-sustained cold-atom oscillator,"
Phys. Rev. A {\bf{97}}, 043406 (2018).


\bibitem{barak2005}
R. Barak and Y. Ben-Aryeh,
“Non-orthogonal positive operator valued measure phase distributions of one- and two-mode electromagnetic fields,"
J. Opt. B: Quantum Semiclass. Opt. {\bf{7}}, 123 (2005).


\bibitem{zhirov2006}
O. V. Zhirov and D. L. Shepelyansky,
“Quantum synchronization,"
Eur. Phys. J. D {\bf{38}}, 375 (2006).














\bibitem{goychuk2006}
I. Goychuk, J. Casado-Pascual, M. Morillo, J. Lehmann, and P. H\"anggi,
“Quantum Stochastic Synchronization,"
Phys. Rev. Lett. {\bf{97}}, 210601 (2006).

\bibitem{zhirov2008}
O. V. Zhirov and D. L. Shepelyansky,
“Synchronization and Bistability of a Qubit Coupled to a Driven Dissipative Oscillator,"
Phys. Rev. Lett. {\bf{100}}, 014101 (2008).

\bibitem{xu2014}
M. Xu, D. A. Tieri, E. C. Fine, J. K. Thompson, and M. J. Holland,
“Synchronization of Two Ensembles of Atoms,"
Phys. Rev. Lett. {\bf{113}}, 154101 (2014).

\bibitem{giorgi2013}
G. L. Giorgi, F. Plastina, G. Francica, and R. Zambrini,
“Spontaneous synchronization and quantum correlation dynamics of open spin systems,"
Phys. Rev. A {\bf{88}}, 042115 (2013).


\bibitem{roulet2018}
A. Roulet and C. Bruder,
“Synchronizing the Smallest Possible System,"
Phys. Rev. Lett. {\bf{121}}, 053601 (2018).

\bibitem{roulet2018a}
A. Roulet and C. Bruder,
“Quantum Synchronization and Entanglement Generation,"
Phys. Rev. Lett. {\bf{121}}, 063601 (2018).

\bibitem{tan2022}
R. Tan, C. Bruder, and M. Koppenh\"ofer,
“Half-integer vs. integer effects in quantum synchronization of spin systems,"
Quantum {\bf{6}}, 885 (2022).












\end{thebibliography}
\end{document}